\documentclass[aps,pra,showpacs,footinbib,superscriptaddress,twocolumn,longbibliography,10pt]{revtex4-2}

\usepackage{appendix}
\usepackage{amsfonts}
\usepackage{amsmath}
\usepackage{txfonts}
\usepackage{amssymb}
\usepackage{amsbsy} 
\usepackage{graphicx}
\usepackage{color}
\usepackage{mathdots} 
\usepackage{braket} 
\usepackage{lipsum}
\usepackage{hyperref}
\usepackage{float}
\usepackage[normalem]{ulem}
\usepackage[dvipsnames]{xcolor}
\hypersetup{colorlinks=true, linkcolor=blue, anchorcolor=blue, citecolor=blue,urlcolor=blue}
\usepackage{graphicx}  
\usepackage{hyperref}

\def\Re{\rm{Re}}
\def\Im{\rm{Im}}
\def\be{\begin{equation}} \def\ee{\end{equation}}
\def\bea{\begin{eqnarray}} \def\eea{\end{eqnarray}}
\def\nn{\nonumber}

\def\ra{\rangle}
\def\la{\langle}

\usepackage{titlesec}
\titleclass{\subsubsubsection}{straight}[\subsubsection]
\newcounter{subsubsubsection}[subsubsection]
\renewcommand\thesubsubsubsection{\thesubsubsection.\arabic{subsubsubsection}}
\titleformat{\subsubsubsection}
  {\normalfont\normalsize\itshape}{\thesubsubsubsection}{1em}{}
\titlespacing*{\subsubsubsection}{0pt}{1.5ex plus .2ex minus .2ex}{0.5ex plus .2ex}

\titleclass{\subsubsubsubsection}{straight}[\subsubsubsection]
\newcounter{subsubsubsubsection}[subsubsubsection]
\renewcommand\thesubsubsubsubsection{\thesubsubsubsection.\arabic{subsubsubsubsection}}
\titleformat{\subsubsubsubsection}
  {\normalfont\normalsize\itshape}{\thesubsubsubsubsection}{1em}{}
\titlespacing*{\subsubsubsubsection}{0pt}{1.25ex plus .2ex minus .2ex}{0.5ex plus .2ex}

\begin{document}

\title{Topologically protected negative entanglement}

\author{Wen-Tan Xue}
\email{wentanx@nus.edu.sg}
\affiliation{ Department of Physics, National University of Singapore, Singapore 117542}

\author{Ching Hua Lee}
\email{phylch@nus.edu.sg}
\affiliation{ Department of Physics, National University of Singapore, Singapore 117542}

\begin{abstract}
	The entanglement entropy encodes fundamental characteristics of quantum many-body systems, and is particularly subtle in non-Hermitian settings where eigenstates generically become non-orthogonal. In this work, we find that negative biorthogonal entanglement generically arises from topologically protected non-orthogonal edge states in free fermion systems, especially for flat-band edge states. Departing from previous literature which associated negative entanglement with exceptional gapless points, we show that robustly negative entanglement can still occur in gapped systems.
    Gapless 2D flat-band edge states, however, exhibit novel $S_A\sim -\frac1{2}L_y^2\log L$ entanglement behavior which scales quadratically with the transverse dimension $L_y$, independent of system parameters. This dramatically negative scaling can be traced to a new mechanism known as non-Hermitian critical skin compression (nHCSC), where topological and skin localization in one direction produces a hierarchy of extensively many probability non-conserving entanglement eigenstates across a cut in another direction. Our discovery sheds light on new avenues where topology interplays with criticality and non-Hermitian localization, unrelated to traditional notions of topological entanglement entropy. This topologically protected negative entanglement also manifests in the second R\'enyi entropy, which can be measured through SWAP operator expectation values. 
\end{abstract}
\maketitle

The entanglement entropy plays a crucial role in unveiling fundamental insights into the locality of quantum information. For instance, by scaling either according to the volume or area~\cite{Gioev2006,Flammia2009,area2010}, the entanglement entropy reveals whether quantum correlations pervade the entire system or remain localized. Intriguingly, numerous studies~\cite{kitaevTopo,HaldaneTopo,XiaoGang} have suggested that the presence of topological order can also be encoded in the entanglement entropy, as revealed by the presence of an additional constant term~\cite{kitaev2003, Hamma2005} or discontinuities in the scaling relation. 

In this work, we extend the study of entanglement entropy into the non-Hermitian regime~\cite{ZW2018,Lee2019Anatomy,SFchiral,Rijia2023,Jiang2019,Kai2020,Murakami2019,kawabata2019,Amoeba2024,ZSY2020,LLH2021,Sun2021,universal2022,cluster2022,Jiang2023,Tai2023,shen2023,Okuma2023,liu2024localization,lei2024activating}, where a primary feature is that the eigenstates of the Hamiltonian $H$ are generically non-orthogonal. To maintain orthogonality and preserve probabilistic interpretation of quantum mechanics, we employ a biorthogonal basis of left and right eigenstates--i.e., $H=\sum_m E_m\ket{\psi_m^R}\bra{\psi_m^L}$ with $\bra{\psi^L_m}\psi^R_l\rangle=\delta_{ml}$~\cite{Brody2014,Kunst2018,Berry2004,Gong2018,Moiseyev2011}. Within this biorthogonal framework, recent studies have revealed that both bipartite entanglement entropy and R\'enyi entropy can manifest unexpected negative values~\cite{Shinsei2020,EBstate,SciPost2022,fossati2023symmetry,rottoli2024entanglement}, attributable to the presence of geometric defectiveness at exceptional points (EPs)~\cite{EBstate,SciPost2022,fossati2023symmetry,rottoli2024entanglement}. Building upon these insights, we uncover a new mechanism by which topology in non-Hermitian systems substantially influence entanglement entropy behavior. Specifically, we show that certain topological boundary states can exert a strongly non-local influence on the dominant entanglement behavior of the \emph{entire} system, leading to topologically protected negative free-fermion entanglement entropy.

Most notably, in our non-Hermitian model featuring flat-band edge states, we uncover an unconventional negative 
entanglement scaling, given by $S_A\sim-\frac1{2}L_y^2\log L$, where $L$ and $L_y$ are the system dimensions normal and parallel to the entanglement cut, respectively. When the aspect-ratio $L_y/L$ is held fixed, this scaling manifests as a super-volume-law behavior. This 
unconventional scaling reflects a novel quantum correlation structure that is fundamentally distinct from those observed in area-law and volume-law systems, and has rarely been reported in previous studies~\cite{Clerk2024}. The enigmatic $-L_y^2$ scaling dependence arises not just due to the enhanced non-orthogonality of the states due to flatness of the band, but also the extensively many probability-nonconserving entanglement eigenstates that emerge due to the band criticality -- in a new mechanism that we dub non-Hermitian critical skin compression (nHCSC).
 
Another key discovery of this work is that the presence of an EP is not strictly a prerequisite for observing negative entanglement entropy values -- instead, substantial non-orthogonality among the right eigenstates suffices, and spectacularly so when the non-orthogonality is enforced by the flat-band edge states. We investigate two 2D topological non-Hermitian models where the topological edge states in these models demonstrate nearly perfect overlap, while the overlap among bulk states remains minimal. Remarkably, this enables topological nontrivialness to be strategically employed to switch the negative entanglement entropy on or off. 
  
\section{Results}
We present a representative 2D system in which nearly flat topological edge bands give rise to negative entanglement entropy that scales quadratically with $L_y$ in an unconventional manner. We impose a twisted~\footnote{Such that the discrete momentum takes values of $k=\pi/L, 3\pi/L,...$, avoiding the singular $k=0$ point exactly. } cylindrical geometry: the $y$ direction hosts $L_y$ unit cells with open boundaries to support edge state, while the circumferential direction hosts $L$ unit cells and admits a good quantum number $k$ (for clarity denoted simply as $k$ and $L$ instead of $k_x$ and $L_x$).  To realize such flat bands across an extended range of $k$, it suffices to consider a minimal 2-component~\footnote{2 topological bands are sufficient, since the NHSE pushes both of them onto the same boundary and makes them overlap significantly.} Hamiltonian, 
\be
H(k,k_y)=\begin{pmatrix}0&te^{-ik_y}+a_0\\te^{ik_y}+(b_0-\cos k)^B & 0\end{pmatrix},
\label{Hssh}\ee
with asymmetric $k$-dependent off-diagonal hoppings. Here $a_0,b_0$ and hopping distance $B>0$ are real, ensuring that $S_A$ remains real (see Methods, Sec. \hyperref[subsec:C2]{C.2}). Under OBCs along $y$, Eq~\eqref{Hssh} reduces to a non-Hermitian SSH chain, which has been well studied via the generalized Brillouin zone (GBZ) method~\cite{ZW2018,Murakami2019}. The system exhibits a real energy spectrum due to non-Bloch PT-symmetry~\cite{Bender2002,Ruter2010,Ganainy2018,yang2023percolation}, while the eigenstates, including edge states, accumulate at one boundary via the non-Hermitian skin effect (NHSE) with skin depth $-2/\log[(b_0-\cos k)^B/a_0]$. The topological non-trivial region follows from the GBZ winding number is
\be
|a_0(b_0-\cos k)^B|\le t^2,
\label{topoCondition}
\ee
which hosts almost-flat topological edge bands, as shown in Fig.~\ref{b0ne1Ek}(b).

The unconventional entanglement entropy scaling we report is tightly linked to these edge bands--specifically their gap and degree of flatness. Due to finite-size effect, the edge bands are \emph{not necessarily} gapless for any $L_y$, resulting in a dependence of band flatness--and hence $S_A$--on $L_y$. Using Schur's determinant identity on the real-space Hamiltonian $[H_\text{y-OBC}(k)]_{y_1,y_2}=(2\pi)^{-1}\int e^{ik_y(y_1-y_2)}H(k,k_y)\,dk_y$, as elaborated in Methods, Sec. \hyperref[subsec:B1]{B.1}, we obtain 
\bea
\det[H_\text{y-OBC}(k)]=[a_0(b_0-\cos k)^B]^{L_y},
\label{det}
\eea
so that the edge-mode energies $E_{e_1}(k), E_{e_2}(k)$ satisfy:
\bea
E_{e_1}(k)E_{e_2}(k)\approx \left[\frac{a_0(b_0-\cos k)^B}{t^2}\right]^{L_y}.
\label{topo-gap}
\eea
This indicates that the edge-state gap $\Delta=2|E_{e_1}| \sim (\text{Const.})^{-BL_y}$ decreases exponentially with $L_y$, approaching zero as $L_y\rightarrow \infty$ without ever exactly closing. Consequently, the flatness of the edge bands--and thus the scaling of $S_A$--is 
exponentially sensitive to $BL_y$. Perfect gap closure occurs
only for $|b_0|\leq 1$, a case we analyze separately below.

\subsection{Negative entanglement from eigenstate non-orthogonality}
To understand how these band properties relate to the observed entanglement features, we first introduce how negative entanglement can emerge from eigenstate non-orthogonality in non-Hermitian systems. In the non-Hermitian context, the density operator that preserves its role as a probabilistic weight is the biorthogonal density matrix $\rho=\ket{\Psi^R}\bra{\Psi^L}$,
where
\be
\ket{\Psi^R}=\prod_{m\in occ}\left(\psi_m^R\right)^\dagger\ket{0},\quad \ket{\Psi^L}=\prod_{m\in occ}\left(\psi_m^L\right)^\dagger\ket{0}
\ee
are the right and left many-body ground states created by bifermionic creation operators 
 $\left(\psi_m^R\right)^\dagger, \left(\psi_m^L\right)^\dagger$ satisfying $\{\psi^L_{m},(\psi_l^R)^\dagger
\}=\delta_{ml}$, such that $\langle \Psi^L\ket{\Psi^R}=1$, even if $\la \Psi^R\ket{\Psi^R}\ne 1,  \langle\Psi^L\ket{\Psi^L}\ne 1$. We specialize to free boson and fermion systems, where the ground state and thermal states are Gaussian states. As such, all correlation functions adhere to Wick's theorem, and the reduced density matrix $\rho_A$ (obtained by tracing out the degrees of freedom in the complementary region $A^c$), can be fully expressed in terms of two-point correlation function within the entanglement subregion $A$ 
~\cite{Ingo2003}. 
The correlation matrix corresponds to the transpose of the projector matrix $P$, $\la c_{x_2,y_2,s_2}^\dagger c_{x_1,y_1,s_1}\ra=\bra{x_1,y_1,s_1}P\ket{x_2,y_2,s_2}$,
where $x,y$ denote lattice sites, $s_{1,2}=+,-$ label sublattice indices, and $P=\sum_{m\in occ}\ket{\psi_m^R}\bra{\psi_m^L}$ projects onto the occupied bands. Thus, the entanglement entropy defined by $\rho_A$ can be directly computed from $P$ restricted to subregion $A$. This restriction is equivalent to applying the a real-space projector $\Gamma_A=\sum_{(x,y,s)\in A}\ket{x,y,s}\bra{x,y,s}$ to $P$, which yields the truncated band projector
\be
\bar{P}=\Gamma_AP\,\Gamma_A=\sum_{m\in occ}\Gamma_A\ket{\psi_m^R}\bra{\psi_m^L}\Gamma_A=\sum_{m\in occ}\ket{\psi_{mA}^R}\bra{\psi_{mA}^L}.
\label{barP}
\ee

Crucially, this $\bar P$ operator contains complete information about the $n$th-order R\'enyi entropy for free fermions~\footnote{An analogous expression also holds for free bosons: $ S_{A,boson}^{(n)}=\frac{1}{n-1}\text{Tr}\Big[\log\Big(\bar{P}^n-(\bar{P}-I)^n\Big)\Big]$.}
 \bea
 S_{A}^{(n)}=\frac{\log\text{Tr}(\rho_A^n)}{1-n}&=&\frac{1}{1-n}\text{Tr}\left[\log\left(\bar{P}^n+(I-\bar{P})^n\right)\right]\nn\\
 &=&\frac{1}{1-n}\sum_i\log\left(p_i^n+(1-p_i)^n\right),\quad
 \label{Sn}
 \eea
which, in the limit of $n\rightarrow 1$, yields the von Neumann entropy
 \bea
  S_A&=&-\text{Tr}\rho_A\log \rho_A=-\text{Tr}\left[\bar{P}\log\bar{P}+(I-\bar{P})\log(I-\bar{P})\right]\nn\\
  &=&\sum_{p_i}-p_i\log(p_i)-(1-p_i)\log(1-p_i),
  \label{pi}
 \eea
where $I$ is the identity matrix and $p_i$ are the eigenvalues of $\bar{P}$. Physically, each $p_i$ represents an occupation probability restricted to subregion $A$, and is real and bounded within $[0,1]$ for Hermitian Hamiltonian. In non-Hermitian settings, however, $p_i$ can become complex and take values beyond $[0,1]$, with magnitudes $|p_i|\gg 1$. This occurs due to the non-conservation of probability across the subregion boundary, as we shall demonstrate. Substituting such large $|p_i|$ into Eqs~\eqref{Sn} and \eqref{pi} can result in an unexpected negative entanglement entropy. 
In general, both R\'enyi entropies $S_A^{(n>1)}$ and von Neumann entropy $S_A$ calculated from complex $p_i$ can be complex. An important exception arises for PT-symmetric Hamiltonians, for which $S_A^{(n)}$ remains real (see Methods, Sec. \hyperref[subsec:C2]{C.2} for details).
The flat-band edge-state model [discussed in Eq.~\eqref{Hssh}] serves as a paradigmatic example.
For more general cases, such as the exceptional crossing model in Eq~\eqref{4bandHk}, where $\text{Im}(S_A^{(n)})\ne 0$, we show both the real and imaginary parts of the entropies, even though the the real part can be more directly measured.

\begin{figure*}
\centering
\includegraphics[width=18cm, height=10cm]{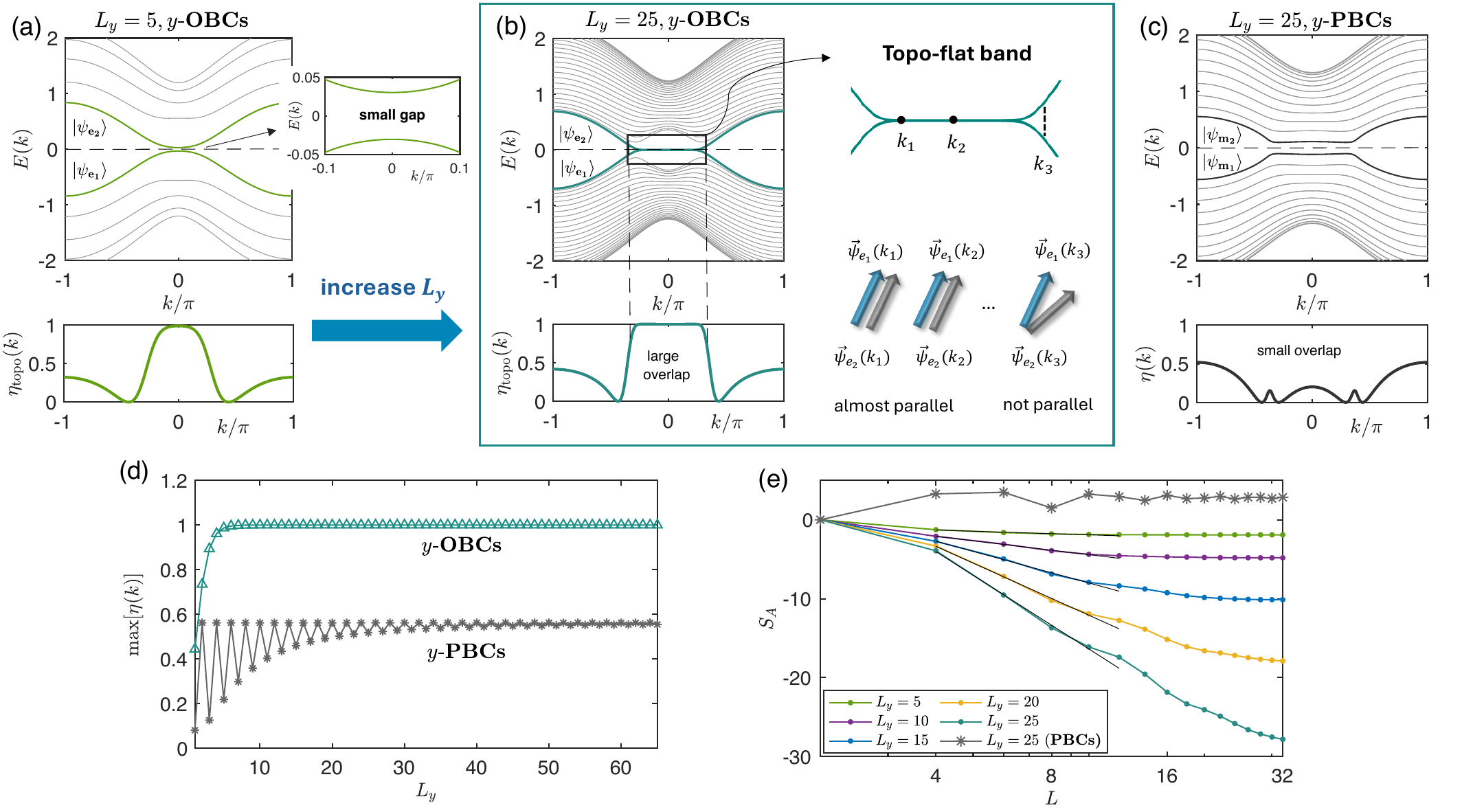}
\caption{Robustly negative entanglement entropy from the gapped flat-band edge states of our 2-component  Hamiltonian (Eq~\eqref{Hssh}), for parameters $B=1, t=0.8, a_0=1$ (with $b_0=1.2\ne 1$ to open up the gap).
(a) For small cylinder length $L_y=5$ and $y$-direction OBCs, the energy spectrum $E(k)$ of the topological edge states $\ket{\psi_{e_1}},\ket{\psi_{e_2}}$ (bolded) exhibits a small but visible gap, but their overlap factor $\eta_\text{topo}(k)$ already approaches unity. (b) Upon increasing $L_y$ to 25, two edge bands with an exponentially small gap is observed within the non-trivial regime prescribed by Eq~\eqref{topoCondition}, with $\eta_\text{topo}(k)\approx 1$ extremely closely. (c) With $y$-PBCs, the midgap flat band disappears and $\eta(k)$ deviates markedly from unity, even though the gap is still small. (d) For $y$-OBCs but not $y$-PBCs, the overlap $\eta_\text{topo}(k)$ saturates very close to unity once $L_y\sim 10^1$. 
(e) The entanglement entropy scaling behavior $S_A$ for different $L_y$. Notably, as $L_y$ increases, $S_A$ decreases with $\log L$ more rapidly as $S_A\sim -(\kappa L_y+\xi)\log L$, with $\kappa\approx0.6633$, $\xi\approx -4.1817$ according to obtained from numerical fitting (black). It also saturates at $S_{min}\sim -L_y$ when $L\gtrapprox L_y$.  
}
\label{b0ne1Ek}
\end{figure*}

Below, we show that mathematically, it suffices to have strong eigenstate overlap in order to have large $|p_i|$, which in turn results in negative R\'enyi and entanglement entropy. For a pair of non-orthogonal right eigenstates $\ket{\psi_m^R}$ and $\ket{\psi_{l}^R}$, their normalized squared overlap~\footnote{$\eta$ can be viewed as a variant of the Petermann factor~\cite{Wiersig2023,YePeng2023,Kerry2020}.}
\be
\eta=\frac{|\bra{\psi_m^R}\psi_{l}^R\ra|^2}{\bra{\psi_m^R}\psi_m^R\ra\bra{\psi_{l}^R}\psi_{l}^R\ra}=\frac{(U^\dagger U)_{m{l}}^2}{(U^\dagger U)_{mm}(U^\dagger U)_{ll}}
\ne 0
\label{Eta}
\ee
does not vanish. Here we have introduced the matrix $U$ whose elements are the real space components of the right eigenstates i.e. $\ket{\psi_{l}^R}=\sum_i U_{i{l}}\ket{i}$, such that the corresponding matrix for the left eigenstates is $U^{-1}$ i.e. $\bra{\psi_m^L}=\sum_i \left(U^{-1}\right)_{mi}\bra{i}$. In the extreme limit where the two eigenstates become parallel, $\eta \rightarrow 1$ and the rank of $U$ becomes lower than the dimension of the space of occupied states. This leads to the vanishing of $\text{Det}(U)$ and crucially forces $U^{-1}$ to acquire very large matrix elements. From
\begin{align}
\sum_{p_i}p_i^2=\text{Tr}(\bar{P}^2)
&=\sum_{m,{l}\in occ}\bra{\psi_{m}^L}\Gamma_A\ket{\psi_{l}^R}\bra{\psi_{l}^L}\Gamma_A\ket{\psi_{m}^R}\notag\\
&=\sum_{m,{l}\in occ}(U^{-1}\Gamma_AU)_{m{l}}(U^{-1}\Gamma_AU)_{{l}m}
\label{eq6}
\end{align}
where we have used $\Gamma_A^2=\Gamma_A$, we deduce that at least one of the absolute values of $p_i$ must also have become very large, since the divergent elements in $U^{-1}$ do not in general cancel off with the small elements in $\Gamma_AU$ except in the case of vanishing entanglement cut $\Gamma_A=I$. However, we stress that even when $U$ is still full-rank with non-defective eigenspace, $\eta$ can already be extremely close to unity and contribute to negative entanglement.
Additionally, the pair of states $\ket{\psi_m^R}$ and $\ket{\psi_l^R}$ are not arbitrarily chosen neighboring bands. Instead, we focus specifically on the two bands that straddle the Fermi level, as shown in Figs.~\ref{b0ne1Ek}(a-c). Intuitively, at zero-temperature, the system is dominated by states near $E_F$, so these are the physically relevant ones. The other reason is, if we redefine $P$ as a projector onto the unoccupied bands, the eigenvalues of $\bar{P}$ transform as $p_i'=1-p_i$. Thus, whenever a divergent $p_i$ appears (leading to negative entanglement entropy), the corresponding $p_i'$ also diverges. This indicates that the emergence of negative EE requires the divergence to be shared between occupied and unoccupied sectors. Therefore, we focus on the overlap between the two states adjacent to $E_F=0$: when these are topological edge states ($\ket{\psi^R_{e_1}}$ and $\ket{\psi^R_{e_2}}$, hereafter denoted simply as $\ket{\psi_{e_1,e_2}}$), we label the overlap as $\eta_\text{topo}(k)$; when they are bulk states ($\ket{\psi_{m_1}}$ and $\ket{\psi_{m_2}}$), we use the usual $\eta(k)$.

\subsection{Unconventional entanglement scaling induced by flat-band edge states} 

To study the entanglement properties of the model in Eq~\eqref{Hssh}, we take half of the $x$ direction, i.e. $L/2$ unit cells, as the entanglement subregion $A: x\in [1,L/2], y\in [1,L_y]$.
The corresponding projector matrix is $\bra{x_1,y_1,s_1}P\ket{x_2,y_2,s_2}=L^{-1}\sum_ke^{ik(x_1-x_2)}\left[P(k)\right]^{s_1,s_2}_{y_1,y_2}$, where $(x_{1,2},y_{1,2})\in A$ and $P(k)=\sum_{m\in occ}\ket{\psi^R_m(k)}\bra{\psi^L_m(k)}$ is a $2L_y\times 2L_y$ projector onto the occupied lower half bands with $\text{Re}[E(k)]<E_F=0$ unless otherwise stated.
The key idea of our flat-band edge-state model and the resulting unconventional entanglement scaling is that, the NHSE localizes all states towards a common boundary, such that the states would exhibit extremely high overlap if they are furthermore macroscopically energetically degenerate, as in the flat-band edge states. This conclusion holds even when the flat bands are not strictly gapless or defective (as in case 1 below). In the following, we analyze two representative situations--gapped and gapless edge states--and examine their corresponding EE scaling behavior.

\begin{figure*}
\centering
\includegraphics[width=18cm, height=11.5cm]{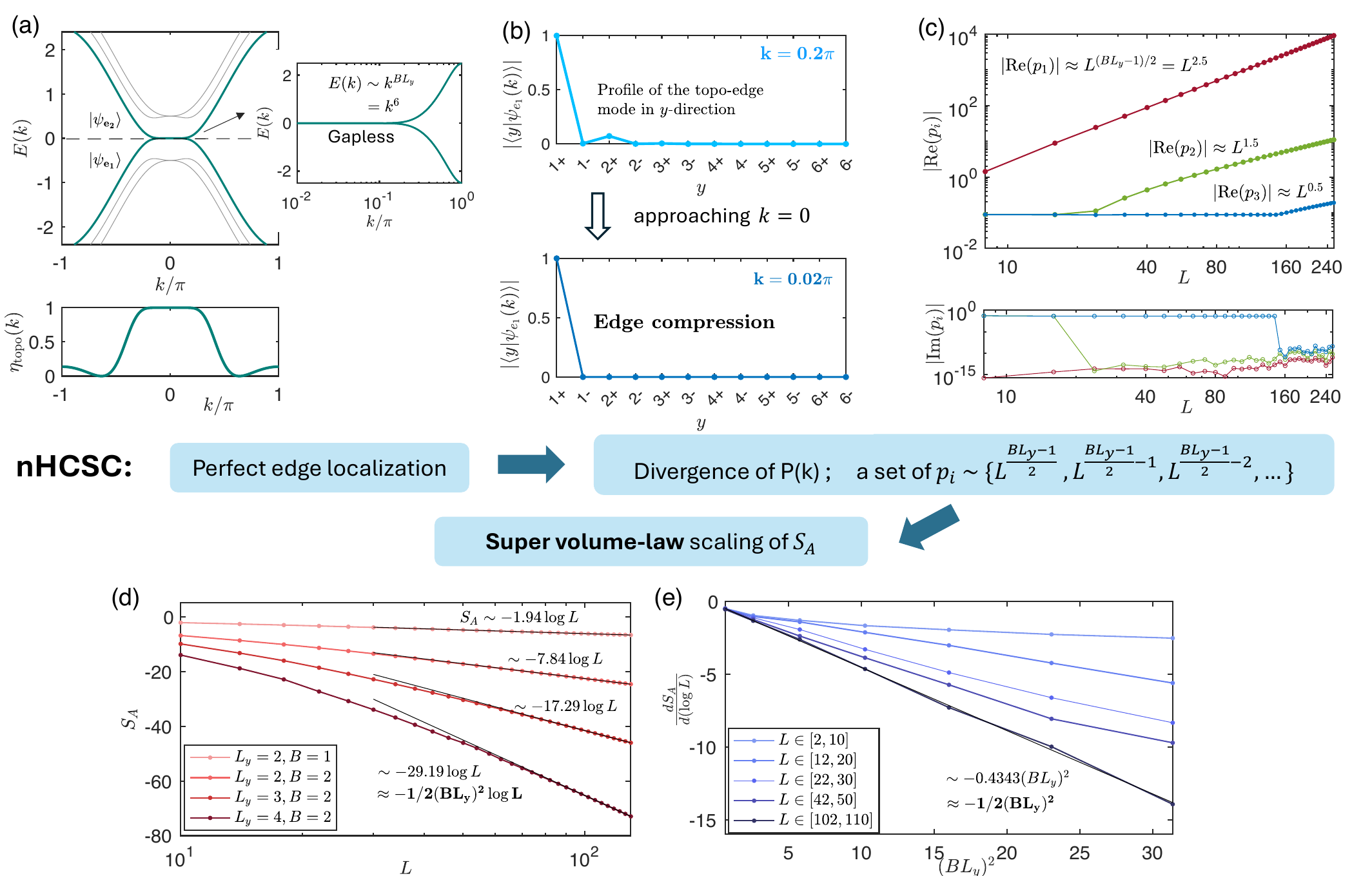}
\caption{Very robust quadratic scaling of the negative entanglement entropy $S_A$ arising from the gapless flat-band edge states of our 2-component Hamiltonian (Eq~\eqref{Hssh}) with parameters $b_0=1, t=0.5, a_0=2$. 
(a) Even for small $L_y=3, B=2$, two nearly flat gapless edge bands (bolded), with dispersion $E_{e_1,e_2}(k)\sim k^{BL_y}$ and overlap $\eta_\text{topo}(k)\approx 1$, emerge around $k=0$ under $y$-OBCs. 
(b) As $k$ approaches $0$, the skin effect ($r(k)\rightarrow 0$) strongly compresses the topological edge mode onto site $1+$, leading to nearly perfect localization.
(c) For this gapless case, occupancy eigenvalues $p_i$ (Eq~\eqref{pii}) of $\bar P$ dramatically exceed the $[0,1]$ interval due to non-Hermitian critical skin compression (nHCSC), with $p_1,p_2,p_3,...$ (red, green, blue...) exhibiting a hierarchy of power-law dependencies with $L$. 
(d) The negative entanglement scaling is accurately approximated by $S_A\approx -\frac1{2}(BL_y)^2\log L$ (Eq~\eqref{SLz2}, black) across different $B,L_y$ combinations for sufficiently large cylinder circumference $L$. 
(e) The coefficient of $\log L$ in the numerical $S_A$, extracted through the gradient of the $dS_A/d(\log L)$ plots (shades of blue),  agrees well with $-\frac1{2}(BL_y)^2$ (Eq~\eqref{SLz2}) when $L\gtrapprox 10^2$. At smaller $L$, the dependence $\propto (BL_y)^2$ still holds for smaller $BL_y$, albeit with a smaller coefficient.}
\label{b0eq1}
\end{figure*}

\noindent\emph{1. Gapped edge states ($b_0> 1$). --}
Even though the edge bands are gapped, they become almost flat and touching as $L_y$ is increased, as depicted in Figs.~\ref{b0ne1Ek}(a,b). At large $L_y$, their gap becomes exponentially small within the topologically non-trivial region given by Eq~(\ref{topo-gap}), where the overlap factor 
$\eta(k)\approx 1$, indicating that $\ket{\psi_{e_1}}$ and $\ket{\psi_{e_2}}$ are nearly parallel. This suggests that states within an extensive continuum of $k$ closely approximate EPs, 
which is unexpected since the Hamiltonian $H(k,k_y)$ in Eq~\eqref{Hssh} does not inherently feature EP crossings. 
As a comparison, for periodic boundary conditions(PBCs) in the $y$ direction [Fig.~\ref{b0ne1Ek}(c)], the flat-band edge states are absent, and the overlap $\eta(k)$ does not approach $1$ even as $L_y$ increased to a large value of $25$ where the (bulk) band gap becomes quite narrow. For $y$-OBCs, even at very small $L_y\approx 5$ number of layers, $\eta(k)$ is already very close to one [Fig.~\ref{b0ne1Ek}(d)]; at larger $L_y$, $\eta(k)$ converges exponentially to 1 despite the system being physically gapped. 

This strong flat-band edge state-induced non-orthogonality ($\eta\approx 1$) is manifested in a strongly negative bipartite entanglement entropy $S_A$. As shown in Fig.~\ref{b0ne1Ek}(e), $S_A$ scales negatively with the cylinder circumference $\log L$, with a gradient that grows with its length $L_y$. From Methods, Sec. \hyperref[subsec:B3]{B.3}, the exact dependence is established as $S_A\sim -({\kappa} L_y+{\xi})\log L$, where $\kappa\approx0.6633$, $\xi\approx -4.1817$ as obtained from numerical fitting. Notably, this linear dependence on $L_y$ does \emph{not} arise trivially because the length of the entanglement cut scales with $L_y$, since it is contributed only by the topological edge modes whose number do not scale extensively with system length. Rather, it arises because the band flatness scales exponentially with $L_y$. That said, for a given $L_y$, the entanglement entropy $S_A$ saturates at negative lower bound (Methods, Sec. \hyperref[subsec:B3]{B.3}) $S_{min}\sim -L_y\log[a_0(b_0-1)^{-B}]$ because the system is ultimately gapped, such that the overlap $\eta$ at $k_1=\pi/L$ (nearest point to $k=0$) does not approach arbitrarily close to 1 with increasing $L$. 
For PBCs, the bulk gap also results in the saturation of $S_A$ at a positive value, as depicted by the starred grey trend in Fig.~\ref{b0ne1Ek}(e). 

\noindent\emph{2. Gapless edge states ($b_0=1$) with unconventional negative entanglement. --} 
Finally, we discuss the most intriguing case where $\det[H_\text{y-OBC}(k)]=0$ at $k=0$ [Eq~\eqref{det}], such that the edge-state gap vanishes exactly [Fig.~\ref{b0eq1}(a)]. Even though its band structure looks superficially similar to the $b_0>1$ case with exponentially small gap [Fig.~\ref{b0ne1Ek}], its entanglement entropy exhibits a surprising 
dependence $S_A\sim  -\frac1{2}B^2L_y^2\log L$, proportional not to the cylinder length $L_y$, but to the \emph{square} of it. While the first power of $L_y$ can be attributed to the exponentially high topological band flatness as before, the additional second power of $L_y$ emerges from an uniquely new 2D phenomenon which we call non-Hermitian critical skin compression (nHCSC).

To understand the nHCSC, let us recall the similarity transformation approach used in 1D NH SSH model (see Sec.~\hyperref[subsec:B2]{B.2} in Methods). In the present 2D case, the asymmetry hopping along the $y$ direction introduces an exponential spatial factor into all eigenstates. For the two topological edge bands (one is occupied with Re$[E_{e_1(k)}]<0$, the other is unoccupied with Re$[E_{e_2(k)}]>0$), the edge states take the asymptotic forms $\bra{\psi_{e_1(\text{or }e_2)}^L(k)}y,s\rangle\sim r(k)^{-y}$ and $\langle y,s\ket{\psi_{e_1(\text{ or }e_2)}^R(k)}\sim r(k)^{y}$ with $r(k)$ defined in Eq~\eqref{rk} (we consider parameters such that $r(
k)<1$). When $r(k)$ is sufficient small, the skin effect becomes strong enough to push all states---including the topological edge states---towards the same cylinder boundary ($y=1$ for right eigenstates and $y=L_y$ for left eigenstates). At the same time, the two edge bands approach each other in energy, such that the right edge states $\ket{\psi^R_{e_1}(k)}$ (occupied) and $\ket{\psi^R_{e_2}(k)}$(unoccupied) acquire a substantially large overlap $\eta_\text{topo}(k)$. According to the discussion near Eq~\eqref{eq6}, this drives $\bra{\psi^L_{e_1,e_2}(k)}$ towards divergence, and projector contribution $\ket{\psi^R_{e_1}(k}\bra{\psi^L_{e_1}(k)}$ comes to dominate $\bar P$ and hence entanglement. At the critical (gapless) point $k=0$ where $r(0)=1-\cos k=0$, the edge state becomes perfectly skin-localized (see Fig.~\ref{b0eq1}(b)) in the $x$-direction along the cylinder edge.

Ordinarily, this perfect edge localization (or ``compression") only leads to irreversible 1D non-Bloch dynamics~\cite{longhi2020non} and singular generalized Brillouin zones ~\cite{Guo2021,Stlhammar2021,Denner2023}. However, in our 2D topological entanglement context, it also causes the occupied band projector
\be
[P(k)]^{s_1,s_2}_{y,y'}\approx \la y,s_1\ket{\psi_{e_1}^R(k)}\bra{\psi_{e_1}^L(k)}y',s_2\ra\sim r(k)^{y-y'}
\ee
to diverge for matrix blocks $y<y'$, with strongest divergence in $[P(k)]_{1,L_y}^{+,-}\sim r(k)^{-L_y}\sim (1-\cos k)^{-BL_y/2}$ (see Sec.\hyperref[subsec:B2]{II B.2}). Notably, the most strongly divergent contribution $\sim k^{-BL_y}$ from $(1-\cos k)^{-BL_y/2}$ does \emph{not} dominate the total negative entanglement; of also substantial significance are the entire set of divergent terms $k^{-BL_y+2},k^{-BL_y+4},k^{-BL_y+6},...$ from the sub-leading terms in the expansion of  $(1-\cos k)^{-BL_y/2}$, as well as other $[P(k)]^{s_1,s_2}_{y,y'}$. Consequently, distinct from ordinary EP crossings~\cite{Shinsei2020,EBstate}, essentially the \emph{entire} set of $\bar P$ eigenvalues $p_i$, diverges with $L$: 
\be 
p_i\approx \text{Re}(p_i)\sim\{L^{\frac{BL_y-1}{2}}, L^{\frac{BL_y-1}{2}-1}, L^{\frac{BL_y-1}{2}-2},...\},
\label{pii}
\ee
where the imaginary parts Im$(p_i)$ remain negligible, as shown in Fig.~\ref{b0eq1}(c). This is the main consequence of nHCSC, which hinges on both the edge compression of the eigenstates and its criticality (vanishing of $r(k)$). The hierarchy of these divergent occupancy eigenvalues is shown in Fig.~\ref{b0eq1}(c): upon closer inspection, subdominant $p_2,p_3\sim L^{\frac{BL_y-1}{2}-1},L^{\frac{BL_y-1}{2}-2}$ eigenvalues (green, blue) are observed in addition to the dominant $p_1\sim L^{\frac{BL_y-1}{2}}$. Summing over them, the total entanglement entropy scales like (Methods, \hyperref[subsec:B2]{B.2})
\bea
S_A &=&-\sum_{p_i}p_i\log p_i + (1-p_i)\log(1-p_i) \nn\\
&\approx &-\left[\frac{BL_y-1}{2}+\left(\frac{BL_y-1}{2}-1\right)+\left(\frac{BL_y-1}{2}-2\right)+...\right]\log L\nn\\
&\approx& -\frac1{2}(BL_y)^2\log L.
\label{SLz2}
\eea
This strongly negative entanglement $S_A$ is plotted in Fig.~\ref{b0eq1}(d) for various $B,L_y$, and can be as low as $-70$ for reasonably large $L_y=4, B=2$. By examining the slope of $S_A$ with respect to the universal $\log L$ factor, it is numerically confirmed in Fig.~\ref{b0eq1}(e) that the quadratic coefficient $-(BL_y)^2/2$ accurately holds for across a wide range of $BL_y$ as long as $L\gtrapprox 10^2$ (even though moderately large $L\sim \mathcal{O}(10)$ suffices when $BL_y$ is also of $\mathcal{O}(10)$).

\subsection{ Prospects for measuring negative entanglement through the second R\'enyi entropy} 
\label{subsec:swap}

As established in subsection \hyperref[subsec:C1]{C.1} of the Methods, the negative entanglement also manifests generically as negative R\'enyi entropy. Below, we outline a scheme for measuring the second R\'enyi entropy, defined in the biorthogonal basis as:
\be
S_A^{(2)}=-\log\text{Tr}\big[(\rho_A)^2\big],
\ee
where the reduced density matrix is given by $\rho_A=\text{Tr}_{A^c}\Big[\ket{\Psi^R}\bra{\Psi^L}\Big]$.

\noindent A known approach~\cite{swapnat,swapPRL,swapQMC} for measuring the second R\'enyi entropy or quantum purity involves the SWAP operator, which exchanges two copies of a quantum state:
\be
\text{SWAP}\ket{\psi_1}\otimes\ket{\psi_2}=\ket{\psi_2}\otimes\ket{\psi_1}.
\ee
A commonly used corollary \cite{Ekert2002} is $\text{Tr}(\text{SWAP}\rho_1\otimes \rho_2)=\text{Tr}(\rho_1\rho_2)$, from which the second R\'enyi entropy can be calculated from the expectation value of the SWAP operator on two-copies of the many body state ~\cite{fermion3} as:
\bea
\bra{\psi}\otimes\bra{\psi} \text{SWAP}_A\ket{\psi}\otimes\ket{\psi}&=&\text{Tr}(\text{SWAP}_A\rho\otimes\rho)\nn\\
&=&\text{Tr}(\rho_A^2)\quad=e^{-S_A^{(2)}},
\eea
where $\text{SWAP}_A$ denotes the application of the 
SWAP operator in subregion $A$. Substituting $\rho=\ket{\Psi^R}\bra{\Psi^L}$, we obtain
\bea
\bra{\Psi^L}\otimes\bra{\Psi^L}\text{SWAP}_A\ket{\Psi^R}\otimes\ket{\Psi^R}&=&e^{-S_A^{(2)}}\nn\\
\big|\bra{\Psi^L}\otimes\bra{\Psi^L}\text{SWAP}_A\ket{\Psi^R}\otimes\ket{\Psi^R}\big|&=&e^{-\text{Re}[S_A^{(2)}]}.
\label{eq17}
\eea
Therefore, to measure the (real part of the) biorthogonal second R\'enyi entropy, $\text{Re}[S_A^{(2)}]$, the most mathematically direct way would be to prepare two copies of the ground state of the Hamiltonian $H$ as $\ket{\Psi^R}\otimes\ket{\Psi^R}$, applying the SWAP operator in subregion $A$, and then measuring its overlap with the ground state $\ket{\Psi^L}\otimes\ket{\Psi^L}$ of another Hamiltonian $H^\dagger$. This approach could potentially be implemented using programmable quantum computers~\cite{Buhrman2001,Linke2018,Wang2024,chang2024}. Post-selection, which has been used in measuring negative conditional entropy~\cite{Salek2014}, will also be useful in simulating the non-Hermiticity~\cite{lin2021real,chen2023high,shen2023}.

Alternatively, it would usually be more practical to measure the physical (not biorthogonal) expectation values of the SWAP operator, either as:
\bea
\langle\text{SWAP}_A\rangle_{RR}&=&\bra{\Psi^R}\otimes\bra{\Psi^R} \text{SWAP}_A\ket{\Psi^R}\otimes\ket{\Psi^R},\nn\\
&\text{or}&\nn\\
\langle\text{SWAP}_A\rangle_{LL}&=&\bra{\Psi^L}\otimes\bra{\Psi^L} \text{SWAP}_A\ket{\Psi^L}\otimes\ket{\Psi^L}.
\eea
Thus, to measure $\big|\bra{\Psi^L}\otimes\bra{\Psi^L}\text{SWAP}_A\ket{\Psi^R}\otimes\ket{\Psi^R}\big|$ as given in Eq~(\ref{eq17}), one feasible strategy is to prepare a superposition state with known amplitudes $c_1,c_2$ in a physical system:
\be
\ket{\Psi}=c_1\ket{\Psi^R}\otimes\ket{\Psi^R}+c_2\ket{\Psi^L}\otimes\ket{\Psi^L},
\ee
and then measure the expectation value $\bra{\Psi}\text{SWAP}_A\ket{\Psi}$. The left-right overlap terms (i.e. Eq~\eqref{eq17}) can subsequently be calculated by subtracting the contributions from $\langle\text{SWAP}_A\rangle_{RR}$ and $\langle\text{SWAP}_A\rangle_{LL}$, which can also be independently measured. Moreover, the OBC spectrum of our model, as specified in Eq.~(9), is purely real, leading to identical energies for $\ket{\Psi^R}$ and $\ket{\Psi^L}$ and thus facilitating the preparation of their superposition. 

Beyond the approach described above, other potentially feasible ways for observing negative entanglement can involve  directly measuring the reduced density matrix to calculate the entanglement entropy through quantum state tomography ~\cite{Tomography2005,Torlai2018,Tomography2021}, as well as measuring the non-local correlations in phononic crystals \cite{Lin2024}, from which the entanglement entropy can be inferred. Related quantum simulations through quantum Monte Carlo approaches~\cite{swapQMC,QMC202206} and ultracold atomic optical lattices~\cite{Goldman2016,Guzik2012,swapnat,swapPRL,LiJiaming2019} can also reveal the associated quantum correlations.

\subsection{Discussion}
We have discovered an unconventional scaling of free fermion entanglement entropy in non-Hermitian, topologically non-trivial systems--particularly those
featuring flat-band edge states. Specifically,
the entropy scales quadratically as $S_A\sim- \frac1{2} L_y^2\log L$, despite the entanglement cut being only of length $L_y$. This highly unconventional scaling goes beyond the well known area and volume law, and to our knowledge, has no analogue in Hermitian systems. We identify the origin of this behavior as a new mechanism we call non-Hermitian critical skin compression (nHCSC), where the criticality of highly degenerate NHSE-compressed topological modes gives rise to an extensive hierarchy of probability non-conserving $\bar P$ eigenstates that gives rise to even stronger negative $S_A$. Moreover, we find that this negative scaling feature relies on substantial eigenstate overlap around the Fermi surface, which is less stringent than the previously suggested requirement of an exceptional crossing~\cite{Shinsei2020}. As such, macroscopically degenerate flatbands resulting from simultaneous topological and non-Hermitian skin localization can lead to strongly negative entanglement entropy, even in the presence of a small gap. Importantly, this negative entanglement also applies to the R\'enyi entropy which can be physically measured as suggested in previous subsection \hyperref[subsec:swap]{C}, placing topology as a potentially practical control knob for probability non-conserving negative entanglement.

\begin{figure*}
\centering
\includegraphics[width=18cm, height=5.5cm]{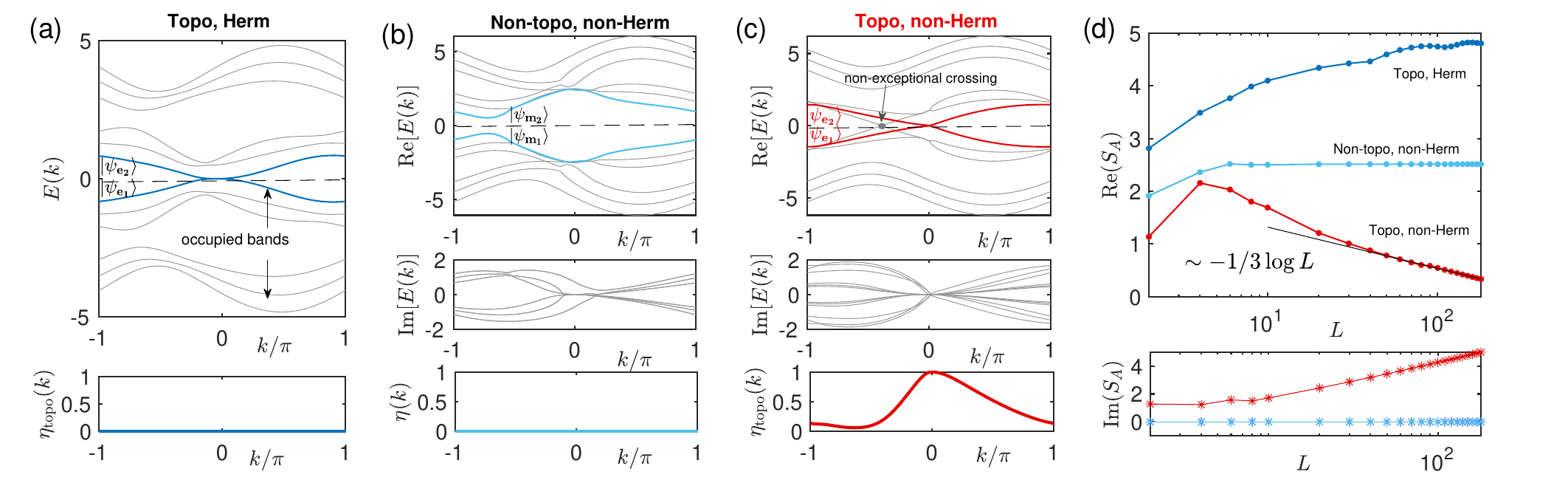}
\caption{Negative entanglement in the 4-band exceptional topological crossing model (Eq~\eqref{4bandHk}) under $y$-OBCs with $L_y=3$. (a) In the topologically non-trivial but Hermitian case ($\alpha=0, M=1.2, \delta=0$), the overlap $\eta_\text{topo}(k)$ [Eq~\eqref{Eta}] of the topological edge states (blue) vanishes rigorously. (b) In the topologically trivial (line-gapped along Re$(E)=0$), non-Hermitian case ($\alpha=0.5\pi,M=3,\delta=2$), $\eta(k)$ of the closest bulk states (light blue) still vanishes essentially. (c) For the non-trivial Chern case ($\alpha=0, M=3, \delta=2$), perfect overlap i.e. $\eta_\text{topo}(k)=1$ is reached where topological edge modes (red) cross. (d) The free fermion entanglement entropy $S_A$ (considering only the real part) for cases (a,b) respectively increases and saturates with system circumference $L$ as expected, but that from the topological exceptional crossing (c) exhibits a new $-\frac1{3}\log L$ scaling. The entanglement subregion is taken to be the half-cylinder with width $L/2$.
}
\label{model1}
\end{figure*}

\section{Methods}

\subsection{A comparative 4-band model with exceptional topological crossing}

\subsubsection{Negative entanglement from exceptional topological crossing} 

In this section, we showcase a 4-band model featuring two topological edge modes that intersect at an exceptional crossing, rather than forming flat bands. We compare its entanglement scaling behavior with that of the flat-band model discussed in the Results,
thereby providing a more comprehensive perspective on topologically protected negative entanglement. Unlike typical topological band crossings~\cite{Hasan2010,Armitage2018} where the topological modes just have to be energetically degenerate, here we require them to also coalesce i.e. become parallel. A candidate model is given by the following 4-band Hamiltonian (see Methods, Sec. \hyperref[subsec:A2]{A.2})
\bea
\mathcal{H}(k,k_y)&=&(\cos k_y-\sin k-M)\tau_x\sigma_0\nn\\ &+&\tau_y(\cos k\sigma_x-\sigma_y+\sin k_y\sigma_z)\nn\\
&+&(\sin \alpha\tau_0+\cos\alpha\tau_x)\sum_{\mu=x,y,z}\sigma_\mu+i\delta\tau_y\sigma_0.
\label{4bandHk}
\eea
where the $\sigma_\mu$ and $\tau_\mu$ Pauli matrices act in spin and sublattice space respectively. The first term controls the band inversion through $M$, the second term represents the spin-orbit coupling which break time-reversal, and the third term introduces a Zeeman field that can also involve sublattice hoppings. The final term, $i\delta\tau_y\sigma_0$, introduces non-Hermiticity through sublattice hopping asymmetry.

In Fig.~\ref{model1}, we present three distinct scenarios corresponding to different parameter combinations, focusing particularly on the overlap $\eta(k)$ between the middle two eigenstates which straddle the Fermi energy $E_F=0$ (dashed line). Open boundary conditions (OBCs) are taken only along the $y$ direction, such that $k$ remains a good quantum number. In Fig.~\ref{model1}(a) with intersecting Hermitian topologicale edge modes (blue), $\eta_\text{topo}(k)=0$ due to the exact orthonormality of Hermitian eigenstates. In Fig.~\ref{model1}(b) which is non-Hermitian ($\delta\neq 0$), $\eta(k)$ remains essentially zero due to the substantial line gap along Re$(E)=0$. However, in the non-Hermitian case with edge modes [Fig.~\ref{model1}(c)], the edge modes (red) cross and coalesce, forming an exceptional point, as reflected by the saturated squared overlap of $\eta_\text{topo}(0)=1$. Only for this exceptional topological case do we see negatively entanglement entropy $\text{Re}(S_A)$ [Fig.~\ref{model1}(d)]; for the previous two gapless and gapped cases of Figs.\ref{model1}(a,b), $\text{Re}(S_A)$ respectively grows/saturates with $L$ as expected from usual conformal field theory~\cite{Calabrese2004,HOLZHEY1994,Hastings_2007}.

Empirically, the 
\be
\Re (S_A)\sim -0.3399\log L\approx \left(\frac{1}{3}-\frac{2}{3}\right)\log L,
\ee
scaling in the exceptional topological case differs from the previously reported $S_A\sim -\frac{2}{3}\log L$ scaling for a linearly dispersive exceptional point~\cite{Shinsei2020,EBstate,SciPost2022}. This discrepancy is attributed to non-exceptional gapless crossing [gray in Fig.~\ref{model1}(c)], which contributes the usual $\frac1{3}\log L$ entanglement. As such, the negative entanglement from exceptional topological crossings can be easily overshadowed by other non-exceptional topological crossings, and is in this sense not necessarily robust~\footnote{This is particularly so if the exceptional dispersion is square-root, as for some generalizations of our model in Methods, subsection \hyperref[subsec:A1]{A.1}.}.

\subsubsection{General form of the Hamiltonian and the EP in its bulk bands}
\label{subsec:A2}

\begin{figure*}
\includegraphics[width=17cm, height=6cm]{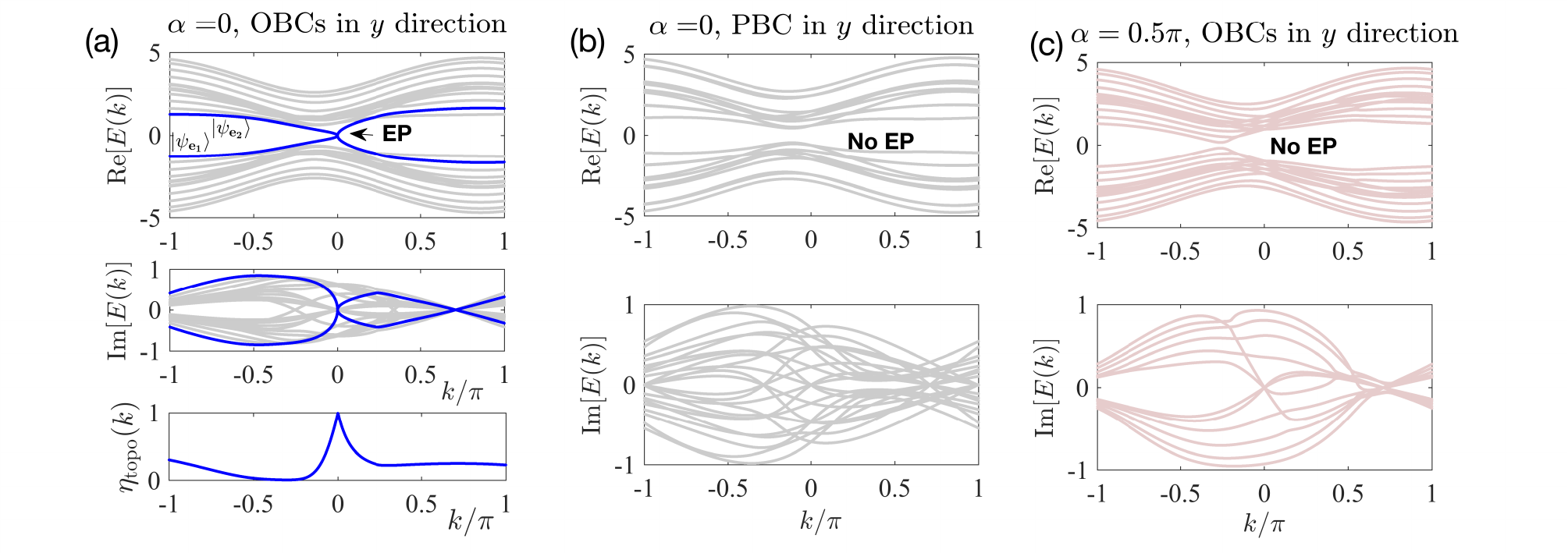}
\caption{Parameter and boundary dependence of EPs in our 4-band model given by Eq.~\eqref{H1General}. (a) For $\alpha=0$ and OBCs in the $y$ direction, an EP occurs at $k=0$. (b) With PBC in the $y$ direction, no EP is observed, implying that the EP arises due to boundary localization. (c) For $\alpha=\pi/2$ with OBCs in the $y$ direction, no EP exists either. Other parameters: $L_y=6, M=3, Z=0.44, \lambda=\delta=1$.}
\label{FigS1}
\end{figure*}

\begin{figure*}
\centering
\includegraphics[width=17cm, height=4.5cm]{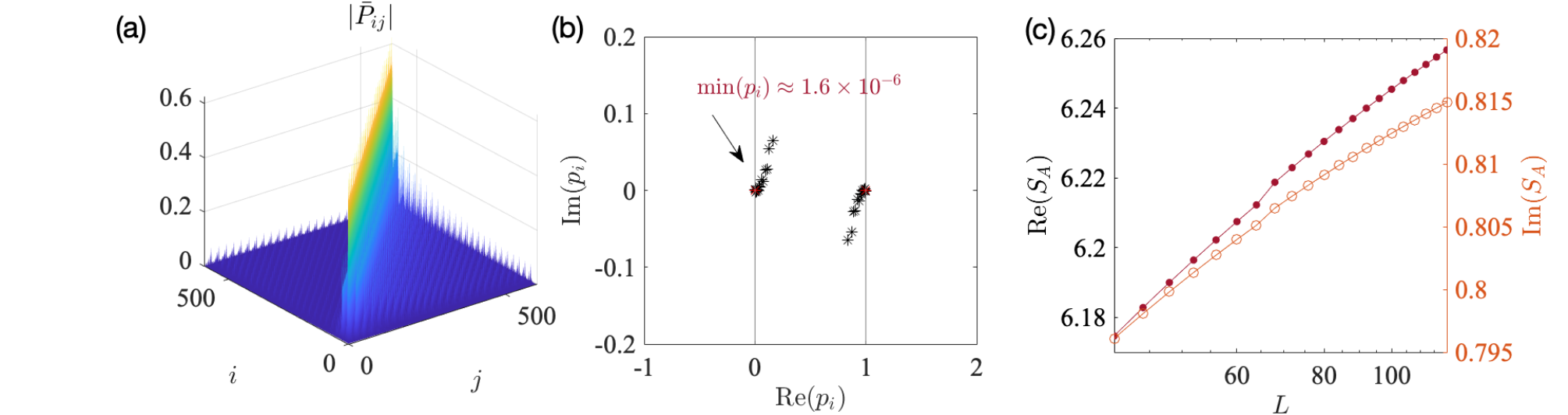}
\caption{Absence of (a) divergent truncated projector $\bar{P}$ matrix elements and (b) eigenvalues outside of $[0,1]$ for an EP with square-root singularities. (c) Consequently, the real part of the entanglement entropy $S_A$ does not exhibit any negativity. Parameters and model are the same as in Fig.~\ref{FigS1}(a), with $x$-direction size $L=50$ and $L_y=6$.}
\label{FigS2}
\end{figure*}

\noindent Here, we show how the 4-band exceptional topological model given by Eq.~(7) of the main text belongs to a more general family of extended exceptional topological models that exists three dimensions. A possible extension is~\cite{Denner2021}:\\
\bea
H(\mathbf{k})&=&\left(\sum_{j=x,y,z}\cos k_j-M\right)\tau_z\sigma_0+\lambda\sum_{j=x,y,z}\sin k_j\tau_x\sigma_j\nn\\
&+&[\sin\alpha\tau_0+\cos\alpha\tau_z](\vec{Z}\cdot\sigma)+i\delta\tau_x\sigma_0
\eea
where $\lambda$ represents the strength of spin-orbit coupling, and $\vec{Z}=(Z,Z,Z)^T$ is the Zeeman field of magnitude $\sqrt{3}Z$ in the $(1,1,1)$ direction. Upon the substitution $k_y=-k_0$ with $k_0=\arcsin(Z)$, and relabeling $k_x$ as $k$, the above reduces to
\bea
\mathcal{H}(k,k_z)&=&(\cos k+\cos k_0+\cos k_z-M)\tau_z\sigma_0+\lambda\big(\sin k\tau_x\sigma_x\nn\\
&+&\sin (-k_0)\tau_x\sigma_y+\sin k_z\tau_x\sigma_z\big)+(\sin \alpha\tau_0+\cos\alpha\tau_z)\vec{Z}\cdot\vec{\sigma}\nn\\
&+&i\delta\tau_x\sigma_0,
\eea

 To simplify the notation, we further relabel $k_z$ as $k_y$, apply a rotation to the Pauli matrices as $\tau_x\rightarrow\tau_y\rightarrow\tau_z\rightarrow\tau_x$ and shift the spectrum by substituting $k\rightarrow k+k_0$, yielding the Hamiltonian:
\bea
H(k,k_y)&=&\big(\cos (k+k_0)+\cos k_0+\cos k_y-M\big)\tau_x\sigma_0\nn\\
&+&\lambda\big(\sin (k+k_0)\tau_y\sigma_x+\sin (-k_0)\tau_y\sigma_y+\sin k_y\tau_y\sigma_z\big)\nn\\
&+&(\sin \alpha\tau_0+\cos\alpha\tau_x)\vec{Z}\cdot\vec{\sigma}+i\delta\tau_y\sigma_0.
\label{H1General}
\eea
Our model in Eq.~(7) of the main text can be viewed as a specific instance of this generalized model that possesses the minimal ingredients of exceptional gapless topological modes, characterized by the parameters $\lambda=1, Z=1$, and $k_0=\arcsin(Z)=\pi/2$.

\subsubsubsection{Effect of EP dispersion on entanglement scaling}

In Figs.~\ref{FigS1} and \ref{FigS2}, we aim to demonstrate that the presence of an exceptional points (EP), whether topologically protected or not, does not necessarily imply the occurrence of a negative EE; the outcome also depends on the energy dispersion around the EP. In Fig.~\ref{FigS1}(a,b), we present a topologically non-trivial configuration, characterized by blue lines that represent the topological edge states for only OBCs and not PBCs. As contrasted with 
the linear dispersion around the EP discussed in the main text, this scenario exhibits a square-root dispersion, $E_{e_1(\text{ or }e_2)}(k)\sim \sqrt{\delta k}$.

Correspondingly, the overlap ${\eta_\text{topo}(k)}$ diminishes rapidly away from the EP $k=0$. It should be noted that in a tight-binding model, the momentum $k$ is discretized, with $k_1=\pi/L$ serving as the closest approximation to $k=0$. Given the rapid decay of $\eta_\text{topo}(k)$ around $k=0$, it follows that $\eta_\text{topo}(k_1)$ does not approach 1. Consequently, as observed in Fig.~\ref{FigS2}(a), $\bar{P}$ exhibits only short-range hoppings, with its eigenvalues $p_i$ almost all located within the range $[0,1]$ (Fig.~\ref{FigS2}(b)) just like for an ordinary non-exceptional model (other than the fact that $p_i$ possess imaginary parts). Furthermore, the entanglement entropy remains positive and increases steadily with system size $L$, as shown in Fig.~\ref{FigS2}(c). Evidently, this scenario with $E_{e_{1,2}}(k)\sim \sqrt{\delta k}$ dispersion does not exhibit negative entanglement, suggesting that the dispersion around the EP crucially  
affects whether the entanglement entropy becomes negative. We will elaborate on how this dispersion can be tuned in our model in Methods, subsection \hyperref[subsec:A3]{A.3}.

\begin{figure*}
\centering
\includegraphics[width=17cm, height=4.5cm]{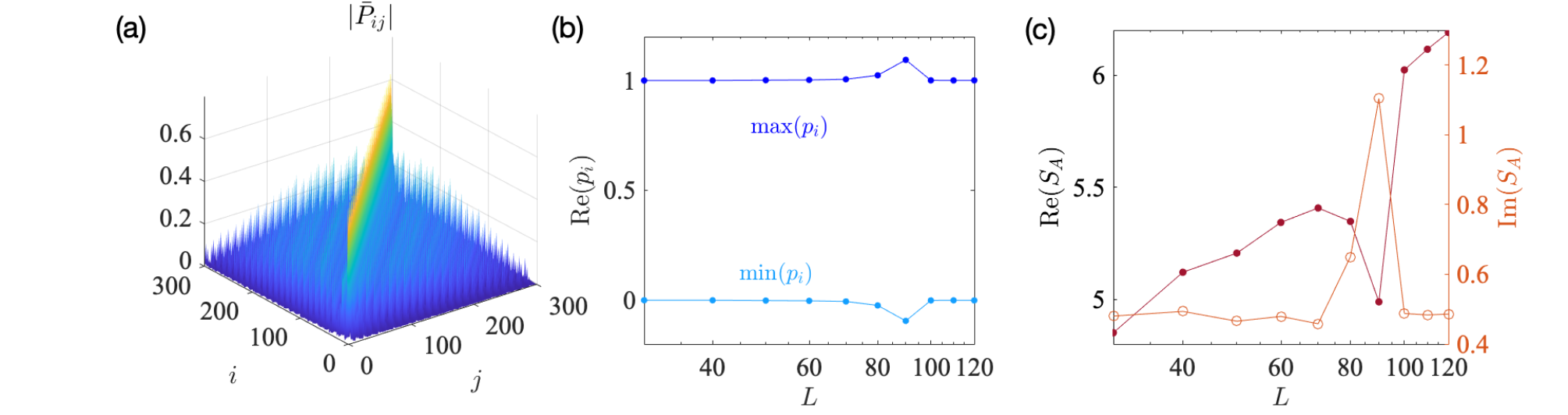}
\caption{Slight negative dip in the entanglement entropy due to bulk EPs. Shown is the model in Eq.~(7) of the main text, with the Fermi Surface fixed at $E_F=-1.1$ and parameters $L_y=3,M=3$ and $\delta=2$. (a) The truncated projector $\bar{P}$  matrix elements for $L=50$ with truncation at $l=L/2=25$. (b) The scaling of the maximum and minimum $p_i$ with system size $L$, which goes out of the $[0,1]$ interval only around $L=90$. (c) The real part of entanglement entropy $S_A$ 
 hence exhibits a slight negative dip at $L\approx 90$. However, this negative dip from the bulk EP is too weak to cause the entanglement to scale negatively as a whole.}
\label{FigS3}
\end{figure*}

\subsubsubsection{Effect of bulk states on entanglement scaling due to topological EPs}

Returning to our model in Eq.~(7) of the main text, we noted that  EPs also exist in the bulk bands around Re$[E(k)]\approx\pm 1.1$. To investigate whether these bulk EPs could similarly induce a negative-valued entanglement entropy, we adjusted the Fermi energy to align with a bulk EP, setting $E_F$ to $1.1$. Observations from Fig.~\ref{FigS3}(a) indicate that the matrix elements of $\bar{P}$ remain minimal, with $p_i$ going beyond $[0,1]$ interval only around $L=90$ (see Fig.~\ref{FigS3}(b)). Consequently, the entanglement entropy exhibits a dip at this system size, albeit only a small dip. This occurs because the bulk EP is located at $k'\approx 0.0106\pi$, rather than at the long-wavelength limit $k=0$. As $L$ increases, $k_1=\pi/L$ will pass through the EP $k'$, inducing a dip in $S_A$. 
However, $S_A$ generally remains positive and continues to increase  with increasing $L$, indicating that the negative scaling observed in Fig.~1(d) of the main text must be predominantly a consequence of the EP in the topological edge bands crossing at $k=0$, rather than of those within the bulk bands.

\subsubsection{ Deriving two types of dispersion relations (linear and square-root) for topological edge bands}
\label{subsec:A3}

For the general form of the Hamiltonian $H(k,k_y)$ as defined in Eq~\eqref{H1General}, when considering OBCs in the $y$-direction with size $L_y$, we obtain:
\be
 H_\text{y-OBC}(k)=\begin{pmatrix} h_0^y &h_+^y &\cdots &0 &0\\h_{-}^y &h_0^y &h_+^y &\cdots&0\\ 0&h_{-}^y&h_0^y&\cdots&0\\
\vdots&\vdots&\ddots&\ddots&\vdots\\0&0&\cdots &h_{-}^y&h_0^y
 \end{pmatrix}_{L_y\times L_y}, 
\ee
with
\be
h_+^y=\begin{pmatrix}0&0&0&0\\0&0&0&1\\1&0&0&0\\0&0&0&0\end{pmatrix},
h_{-}^y=\begin{pmatrix}0&0&1&0\\0&0&0&0\\0&0&0&0\\0&1&0&0\end{pmatrix}, \text{ and }h_0^y(k)=\begin{pmatrix} 0& R\\L & 0\end{pmatrix},
\ee
where
\bea
R=\begin{pmatrix}f+\delta+Z & 2Z-i(Z+\sin (k+k_0)\\i(Z-\sin (k+k_0)) & f+\delta-Z\end{pmatrix},\nn\\ L=\begin{pmatrix}f-\delta+Z & -i(Z-\sin (k+k_0)\\2Z+i(Z+\sin (k+k_0)) & f-\delta-Z\end{pmatrix},
\label{eqS4}
\eea
and $f=\cos (k+k_0)+\cos k_0-M$.
For this $H_\text{y-OBC}(k)$, the topological edge bands exhibit two different types of dispersion relations around the EP: (1) a linear dispersion $E_e(\delta k)\sim \delta k$, as shown in Fig.~\ref{model1}(c), and (2) square-root dispersion $E_e(\delta k)\sim \sqrt{\delta k}$, as depicted in Fig.~\ref{FigS1}(a). 

To explain why the dispersion can exhibit two qualitatively different behavior (i.e., becomes linear when we set $Z=1,\delta=M-Z$ in Eq~\eqref{H1General}), we expand the momentum $k$ around the EP as: $k\rightarrow 0+\delta k$. We shall prove that:
\begin{equation}
\det[H_\text{y-OBC}(\delta k)] \sim
\begin{cases}
    \delta k^2, & \text{with } Z=1, \delta=M-Z\\
    \delta k, & \text{other parameters}
\end{cases}.
\label{eqS5}
\end{equation}
This scaling behavior is directly linked to the dispersion of the topological bands, which are the only bands that become gapless in the spectrum. To elaborate, we know that the determinant is equal to the product of all eigenenergies as: \be
 \det[H_\text{y-OBC}(k)]=E_{e_1}(k)E_{e_2}(k)\prod_n E_n(k),
\ee
where $E_{e_1}(k)$ and $E_{e_2}(k)$ are the two topological edge bands, and the $E_n(k)$s represent bulk bands.
As $\delta k\rightarrow 0$, $\det[H_\text{y-OBC}(\delta k)]$  approaches zero, as do $E_{e_1}(\delta k)$ and $E_{e_2}(\delta k)$ with $|E_{e_{1}}|=|E_{e_2}|$. However, the bulk bands $E_n(\delta k)$ approach finite values. Thus, the topological edge bands exhibit dispersions described by  $E_{e_{1,2}}\sim\sqrt{\det(H_\text{y-OBC})}$. Below, we show that this $\sim\delta k$ when the parameters are set to $Z=1,\delta=M-Z$. For other parameter combinations, $E_{e_{1,2}}$ will behave generically as $\sim\sqrt{\delta k}$.

To derive Eq~\eqref{eqS5}, we first expand $H_\text{y-OBC}$ around $k=0$. Since $k$ appears only in the diagonal element $h_0^y$, we need to expand the $R$ and $L$ matrices [Eq~\eqref{eqS4}].  By substituting $\sin k_0=Z$ into Eq~\eqref{eqS4}, we obtain
\bea \sin (k_0+\delta k)\approx Z+\cos k_0\delta k-\frac{Z}{2}\delta k^2,\nn\\
\cos(k_0+\delta k)\approx \cos k_0-Z\delta k-\frac{\cos k_0}{2}\delta k^2,
\eea
and
\bea
R(\delta k)&=&\begin{pmatrix}R_{11} & R_{12}\\R_{21} & R_{22}\end{pmatrix}\nn\\
&=&\begin{pmatrix}f_0+\delta+Z & 2Z(1-i)\\0 & f_0+\delta-Z\end{pmatrix}+\begin{pmatrix}-Z &-i\cos k_0\\-i\cos k_0 &-Z\end{pmatrix}\delta k\nn\\
&&+\begin{pmatrix}-\frac{1}{2}\cos k_0 & i\frac{Z}{2}\\i\frac{Z}{2}&-\frac{1}{2}\cos k_0 \end{pmatrix}\delta k^2+\mathcal{O}(\delta k^3),
\label{eqR}
\eea
where $f_0=2\cos k_0-M$, and
\bea
L(\delta k)&=&\begin{pmatrix}L_{11} & L_{12}\\L_{21} & L_{22}\end{pmatrix}\nn\\
&=&\begin{pmatrix}f_0-\delta+Z & 0\\2Z(1+i) & f_0-\delta-Z\end{pmatrix}+\begin{pmatrix}-Z &i\cos k_0\\i\cos k_0 &-Z\end{pmatrix}\delta k\nn\\
&&+\begin{pmatrix}-\frac{1}{2}\cos k_0 & -i\frac{Z}{2}\\-i\frac{Z}{2}&-\frac{1}{2}\cos k_0 \end{pmatrix}\delta k^2+\mathcal{O}(\delta k^3).
\label{eqL} 
\eea 
Then, by using Schur’s determinant identity, $\det\begin{pmatrix}A & B\\C & D\end{pmatrix}=\det(D)\det(A-BD^{-1}C)$ ~\cite{Schur2005}, we can calculate $\det[H_\text{y-OBC}]$ as follows (for simplicity, we rewrite $h_0^y$ as $h_0$):
\bea
\det && \begin{pmatrix} h_0 &h_+ &\cdots &0\\h_{-} &h_0 &h_+ &\vdots\\ 0&h_{-}&\ddots& h_+\\
0&\cdots &h_{-}&h_0\end{pmatrix}_{L_y}
=\det(h_0)\det\begin{pmatrix}h_0 &h_+ &\cdots &0\\h_{-} &h_0 &h_+ &\vdots\\ 0&h_{-}&\ddots& h_+\\
0&\cdots &h_{-}&\color{blue}h_0^{(1)}\end{pmatrix}_{L_y-1}\nn\\
&=&\det(h_0)\det(h_0^{(1)})\det\begin{pmatrix} h_0 &h_+ &\cdots &0\\h_{-} &h_0 &h_+ &\vdots\\ 0&h_{-}&\ddots& h_+\\
0&\cdots &h_{-}&\color{blue}h_0^{(2)}\end{pmatrix}_{L_y-2}\nn\\
&=&\det(h_0)\det(h_0^{(1)})\det(h_0^{(2)})\cdots\det(h_0^{(L_y-1)}),
\label{detmulti}
\eea
 in which $h_0^{(1)}= h_0-h_+h_0^{-1}h_-=\begin{pmatrix}0 &R^{(1)}\\L^{(1)}& 0 \end{pmatrix}$, with
 \bea
R^{(1)}&=&R-\begin{pmatrix}0&0\\(R^{-1})_{21}&0\end{pmatrix}, \quad L^{(1)}=L-\begin{pmatrix}0&(L^{-1})_{12}\\0&0\end{pmatrix}
\eea
and $h_0^{(2)}= h_0-h_+[h_0^{(1)}]^{-1}h_-=\begin{pmatrix} 0&R^{(2)}\\L^{(2)}& 0\end{pmatrix}$, with
\bea
R^{(2)}=R-\begin{pmatrix}0&0\\ [(R^{(1)})^{-1}]_{21}&0\end{pmatrix},
 L^{(2)}=L-\begin{pmatrix}0&[(L^{(1)})^{-1}]_{12}\\0&0\end{pmatrix},
 \label{eqS10}
 \eea
 and so forth $h_0^{(n)}=h_0-h_+[h_0^{(n-1)}]^{-1}h_-$. Now, to calculate Eq~\eqref{detmulti}, we need to determine how $\text{det}(h_0^{(n)})$ varies with $\delta k$.

\begin{itemize}
    \item For the case considered in the main text with $ \lambda=Z=1,\delta=M-Z, k_0=\pi/2$, we have
\bea
 R(\delta k)=\begin{pmatrix}-\delta k&2(1-i)+\frac{i}{2}\delta k^2\\\frac{i}{2}\delta k^2 & -2-\delta k\end{pmatrix}+\mathcal{O}(\delta k^3),\nn\\
 L(\delta k)= \begin{pmatrix}2(1-M)-\delta k&-\frac{i}{2}\delta k^2\\ 2(1+i)-\frac{i}{2}\delta k^2 &-2M\end{pmatrix}+\mathcal{O}(\delta k^3),
 \eea
 and $\det(R)\approx 2\delta k, \det(L)\approx-4M(1-M)+2M\delta k$, which gives us:
\be
\det(h_0)=\det(R)\det(L)\approx-8M(1-M)\delta k.
\ee
Importantly, the fact that $\det(R)$ is proportional to $2\delta k$ and vanishes as $\delta k\rightarrow 0$ will turn out to be crucial. According to Eq~\eqref{eqS10}, for the first iteration, $\det(h_0^{(1)})$, can be calculated using $R^{(1)}$ and $L^{(1)}$ 
\bea
R^{(1)}&=&\begin{pmatrix}R_{11}&R_{12}\\R_{21}-(R^{-1})_{21}&R_{22}\end{pmatrix}=\begin{pmatrix}R_{11}&R_{12}\\(1+\frac{1}{\det (R)})R_{21} &R_{22}\end{pmatrix}\nn\\
&\approx&\begin{pmatrix}-\delta k&2(1-i)+\frac{i}{2}\delta k^2\\ \frac{i}{4}\delta k+\frac{i}{2}\delta k^2 & -2-\delta k\end{pmatrix}+\mathcal{O}(\delta k^3),\nn\\
L^{(1)}&=&\begin{pmatrix}L_{11}&L_{12}-(L^{-1})_{12}\\L_{21}&L_{22}\end{pmatrix}=\begin{pmatrix}L_{11}&\left(1-\frac{1}{4M(1-M)}\right)L_{12}\\L_{21} &L_{22}\end{pmatrix}\nn\\
&\approx&\begin{pmatrix}2(1-M)-\delta k&-\frac{i}{2}\left(1-\frac{1}{4M(1-M)}\right)\delta k^2\\ 2(1+i)-\frac{i}{2}\delta k^2 &-2M\end{pmatrix}\nn\\
&&+\mathcal{O}(\delta k^3),
\eea
with $\det(R^{(1)})\approx(2+\frac{1+i}{2})\delta k$, $\det(L^{(1)})\approx -4M(1-M)$ and
\bea
\det(h_0^{(1)})&=&\det(R^{(1)})\det(L^{(1)})\nn\\
&\approx &-4M(1-M)\left(2+\frac{1+i}{2}\right)\delta k.
\eea
\end{itemize}

For the second and subsequent iterations ($n\ge 2$), the elements $R^{(n)}_{11/12/22}$ and $L^{(n)}_{11/21/22}$ remain unchanged. And for element$ \lbrack R^{(n)}\rbrack_{21}$, we have
\bea
 \lbrack R^{(2)}\rbrack_{21}&=& R_{21}- [(R^{(1)})^{-1}]_{21}=R_{21}+\frac{R^{(1)}_{21}}{\det(R^{(1)})}\nn\\
 &\approx& \frac{i}{10+2i}=\text{Const},\nn\\ \det(R^{(2)})&\approx&-\frac{1+i}{5+i}=\text{Const},\nn\\
 \lbrack R^{(n)}\rbrack_{21}&=& R_{21}- [(R^{(n-1)})^{-1}]_{21}=R_{21}+\frac{R^{(n-1)}_{21}}{\det(R^{(n-1)})}\nn\\
 &=&\frac{i}{2}\delta k^2+\frac{\text{Const}}{\text{Const}}\approx \text{Const},\nn\\
 \det(R^{(n)})|_{n\ge2}&\approx &2(1-i)[R^{(n)}]_{21}=\text{Const},
\eea
 and for $ \lbrack L^{(n)}\rbrack_{12}$,
 \bea
\lbrack L^{(2)}\rbrack_{12}&=& L_{12}- [(L^{(1)})^{-1}]_{12}=-\frac{i\delta k^2}{2}+\frac{L_{12}^{(1)}}{\det(L^{(1)})}\nn\\
&\approx&\frac{i\delta k^2}{2}\left(\frac{1-\frac{1}{4M(1-M)}}{4M(1-M)}-1\right),\nn\\
\det(L^{(2)})&\approx &-4M(1-M)=\text{Const},\nn\\
\lbrack L^{(n)}\rbrack_{12}&=& L_{12}+\frac{L_{12}^{(n-1)}}{\det(L^{(n-1)})}\sim\delta k^2,\nn\\
\det(L^{(n)})|_{n\ge2}&\approx&-2(1-M)*2M=-4M(1-M)\nn\\
&=&\text{Const}.
 \eea
Therefore, for $n\ge 2$,  the determinant $\det(h_0^{(n)})$ remains constant, leading to the quadratic dispersion:
\be \det(H_\text{y-OBC})=\det(h_0)\det(h_0^{(1)})\cdots\det(h_0^{(L_y-1)})\sim\delta k^2.\ee
\begin{itemize}
\item 
For most other generic parameter combinations, such as the case shown in
Fig.~\ref{FigS1}(a), the constant term in $\det(R)$ does not vanish i.e., $\det(R)\rightarrow \text{Const}$ as $\delta k\rightarrow 0$. Below we show how that leads to a qualitatively different dispersion. We have  \bea
\det(R)&\approx&(f_0+\delta)^2-Z^2-2Z\big(f_0+\delta-(1+i)\cos k_0\big)\delta k\nn\\
&=&C_0^R+C_1^R\delta k,\nn\\
\det(L)&\approx&(f_0-\delta)^2-Z^2-2Z\big(f_0-\delta+(1-i)\cos k_0\big)\delta k\nn\\
&=&C_0^L+C_1^L\delta k,\nn\\
\det(h_0)&=&\det(R)\det(L)\nn\\
&\approx& C_0^RC_0^L+(C_0^RC_1^L+C_0^LC_1^R)\delta k.
 \eea
where $C_0^{R(L)}$ and $C_1^{R(L)}$ are constants. For the first iteration, we have 
 \bea
R^{(1)}&=&\begin{pmatrix}R_{11}&R_{12}\\R_{21}-(R^{-1})_{21}&R_{22}\end{pmatrix}\approx\begin{pmatrix}R_{11}&R_{12}\\(1+\frac{1}{C_0^R})R_{21}&R_{22}\end{pmatrix},\nn\\
\det(R^{(1)})&\approx &C_0^{(1)R}+C_1^{(1)R}\delta k,\nn\\
L^{(1)}&=&\begin{pmatrix}L_{11}&L_{12}-(L^{-1})_{12}\\L_{21}&L_{22}\end{pmatrix}\approx\begin{pmatrix}L_{11}&(1+\frac{1}{C_0^L})L_{12}\\L_{21}&L_{22}\end{pmatrix},\nn\\
\det(L^{(1)})&\approx& C_0^{(1)L}+C_1^{(1)L}\delta k,\nn\\
\det(h_0^{(1)})&=&\det(R^{(1)})\det(L^{(1)})\nn\\
&\approx & C_0^{(1)R}C_0^{(1)L}+(C_0^{(1)R}C_1^{(1)L}+C_0^{(1)L}C_1^{(1)R})\delta k.
 \eea
 By analogy, for $n\ge 2$, we have 
\bea
\lbrack R^{(n)}\rbrack_{21}&=& R_{21}- [(R^{(n-1)})^{-1}]_{21}=R_{21}+\frac{[R^{(n-1)}]_{21}}{\det(R^{(n-1)})}\nn\\
&\sim& R_{21},
\eea
and similarly, $\lbrack L^{(n)}\rbrack_{12}\sim L_{12}$,
 which ensures that the expansion of $\det(h_0^{(n)})=\det(R^{(n)})\det(L^{(n)})\sim \det(h_0)$ always remains in the form of $C_0^{(n)}+C_1^{(n)}\delta k$. 
Therefore, in this case, we can obtain the linear dispersion of $\det(H_\text{y-OBC})$ as:
 \bea &\det&(H_\text{y-OBC})=\det(h_0)\det(h_0^{(1)})\cdots\det(h_0^{(L_y-1)})\nn\\
&\approx&(C_0+C_1\delta k)*(C_0^{(1)}+C_1^{(1)}\delta k)\cdots(C_0^{(L_y-1)}+C_1^{(L_y-1)}\delta k)\nn\\
&\sim& \delta k,\eea
where the constant term $C_0C_0^{(1)}\cdots C_0^{(L_y-1)}$ goes to zero due to the existence of zero energy modes as shown in Fig.~\ref{FigS1}(a).\\
\end{itemize}
The above approach hinges on the observation that the two types of dispersion are distinguished by whether $\text{det}{(R)}=0$. It works for models where most matrix elements in $h_+^y$ and $h_-^y$ are $0$, which effectively simplifies the calculation of $h_0^{(n)}=h_0-h_+[h_0^{(n-1)}]^{-1}h_-$ in Eq~\eqref{detmulti}. In more general models with hoppings beyond nearest neighbors, Eq~\eqref{detmulti} would need to be extended to handle multiple matrix diagonals, and this iterative approach could become far more complicated.

\subsection{Detailed analysis of the flat-band edge-state model}

\noindent In this section, we analyze the two-band model introduced in Eq.~(9) of the main text 
\be
H(k,k_y)=t\cos k_y \sigma_x +(a_0- t \sin k_y)\sigma_+ + ((b_0-\cos k)^B +t\sin k_y)\sigma_-\,,
\ee
which possesses flat-band edge states that experience the NHSE.

\subsubsection{Exponential scaling of the {edge-state gap $\Delta$} with \texorpdfstring{$L_y$}
{Ly}.} 
\label{subsec:B1}

Here, we show that the {edge-state} gap $\Delta$ of our flat band model exhibits exponential decay with respect to the system size $L_y$ within the topologically non-trivial regime. Considering OBCs in the $y$-direction, as similarly analyzed in the previous section, the Hamiltonian can be expressed as follows:
\bea
H_\text{y-OBC}(k)&=&\begin{pmatrix} h_0^y &h_+^y &\cdots &0 &0\\h_{-}^y &h_0^y &h_+^y &\cdots&0\\ 0&h_{-}^y&h_0^y&\cdots&0\\
\vdots&\vdots&\ddots&\ddots&\vdots\\0&0&\cdots &h_{-}^y&h_0^y\end{pmatrix}_{L_y\times L_y}
\eea
with 
\bea
h_0^y=\begin{pmatrix}0 & a_0\\(b_0-\cos k)^B & 0\end{pmatrix}, \, h_+^y=\begin{pmatrix}0 & 0\\t &0\end{pmatrix},\,h_-^y=\begin{pmatrix}0 & t\\0 &0\end{pmatrix}.
\label{HsshzOBC}
\eea
Specifically, they satisfy the equation $h_+^y(h_0^y)^{-1}h_-^y=0$. Therefore, for this model, we have
\bea
\det(H_\text{y-OBC}(k))&=&\det(h_0^y)\det\begin{pmatrix} h_0^y &h_+^y &\cdots  &0\\h_{-}^y &h_0^y &\cdots&0\\ 
\vdots&\ddots&\ddots&\vdots\\0 &\cdots &h_{-}^y& h_0^y-h_+^y(h_0^y)^{-1}h_-^y\end{pmatrix}\nn\\
&=&\det(h_0^y)\det\begin{pmatrix} h_0^y &h_+^y &\cdots  &0\\h_{-}^y &h_0^y &\cdots&0\\ 
\vdots&\ddots&\ddots&\vdots\\0 &\cdots &h_{-}^y& h_0^y\end{pmatrix}_{(L_y-1)\times (L_y-1)}\nn\\
&=&\left(\det(h_0^y)\right)^{L_y}
\quad =\left(a_0(b_0-\cos k)^B\right)^{L_y}.
\eea
To account for the NHSE experienced by system, we can perform a basis transform to the surrogate Hamiltonian~\cite{lee2020unraveling}. Since $H_\text{y-OBC}$ can be viewed as a non-Hermitian SSH model with $k$-dependent, asymmetric intra-cell hopping amplitudes $a_0$ and $b_0-\cos k$, Therefore, akin to the non-Hermitian SSH model~\cite{ZW2018}, we can apply a similarity transformation to $H_\text{y-OBC}$, 
\bea
H'(k)=Q^{-1}H_\text{y-OBC}(k)Q, Q&=&\text{diag}\{1,r,r,r^2,\cdots r^{L_y-1},r^{L_y-1},r^{L_y}\}, \nn\\
r&=&\sqrt{\left|\frac{(b_0-\cos k)^B}{a_0}\right|},
\label{rk}
\eea
and obtain a Hermitian matrix $H'(k)$:
\bea
H'(k)=\begin{pmatrix}0&te^{-ik_y}+t'\\te^{ik_y}+t' &0\end{pmatrix},
\label{H'k}
\eea
where $t'=\sqrt{a_0(b_0-\cos k)^B}$.
Since a similarity matrix transform does not change the eigenspectrum, $H'(k)$ possesses the same eigenvalues as $H_\text{y-OBC}$, with its bulk eigenvalues given by $E_n=\pm\sqrt{(te^{ik_y}+t')(te^{-ik_y}+t')}$. By substituting these values of $E_n$ into $\det[H_\text{y-OBC}(k)]=E_{e_1}(k)E_{e_2}(k)\prod_n E_n(k)$, we obtain
\bea
\prod_n &E_n(k)&=-\prod_{k_y}(te^{ik_y}+t')(te^{-ik_y}+t')\nn\\
&\sim & t^{2L_y}+c_1t^{2L_y-1}t'+c_2t^{2L_y-2}t'^2+\cdots +c_{L_y}t'^{2L_y},
\label{eqS27}
\eea
where in the RHS, the terms containing $e^{\pm ik_y}$ are incorporated into the coefficients  $c_1,c_2,...,c_{L_y}$. As mentioned in the main text, the topological non-trivial condition is $|a_0(b_0-\cos k)^B|\le t^2$, i.e. $|t'|\le |t|$. Consequently, the leading term in the RHS of the above equation is $t^{2L_y}$. Then using Eq~\eqref{eqS27}, we have
\bea
\det[H_\text{y-OBC}(k)]&=&E_{e_1}(k)E_{e_2}(k)t^{2L_y}\left(1+O(t'/t)\right)\nn\\
&=&\left(a_0(b_0-\cos k)^B\right)^{L_y}.\nn\\
E_{e_1}(k)E_{e_2}(k)&\approx&\frac{\left(a_0(b_0-\cos k)^B\right)^{L_y}}{t^{2L_y}}.
\eea
And because $E_{e_1}=-E_{e_2}$, we ultimately obtain the scaling of $\Delta$ with $L_y$ as follows:
\be
\Delta=2|E_{e_1}|\sim \left(\frac{a_0(b_0-\cos k)^B}{t^2}\right)^{L_y/2}.\
\label{Delta}
\ee

\subsubsection{ Scaling of entanglement entropy \texorpdfstring{$S_A$}{SA} with system sizes \texorpdfstring{$L$}{L} and \texorpdfstring{$L_y$}{Ly} in the case of gapless flat bands}
\label{subsec:B2}
\noindent In this subsection, we derive the scaling relations of $S_A$ as a function of $L_y$ and $L$, specifically Eq~\eqref{SLz2} in the main text:
\be 
S_A\sim -\frac1{2}(BL_y)^2\log L
\ee
for the gapless case of $b_0=1$. This is a highly unusual scaling behavior because $S_A$ scales faster than $L_y$, the number of states (volume) from the $y$-dimension. Below we shall elucidate the origin of this unconventional scaling behavior.
While it may seem that the extra $BL_y$ scaling factor (quadratic vs. linear i.e. volume-law) simply originates from the high-order edge-band dispersion $E(k)\sim k^{BL_y}$, we shall show below that the actual mechanism is more complicated, crucially involves the NHSE.

To establish the scaling relation of $S_A$, we first need to prove that:
\bea
\text{Tr}(\bar{P}^2
)=\sum_i p_i^2
\sim c_1L^{BL_y-1}+c_2L^{BL_y-1-2}+\cdots,
\label{TrP2L}
\eea
where $p_i$ represents the eigenvalues of $\bar{P}$, and $c_1,c_2,...$ are coefficients that are independent of $L$. In other words, that the eigenvalues $p_i$ of $\bar P$ scales like various powers of $L$, up to $L^{BL_y-1}$.

Below, we present a comprehensive derivation of Eq~\eqref{TrP2L}. Under OBCs in the $y$ direction, the NHSE which pushes all right eigenstates in the y-direction towards the same edge, they become highly edge-localized and hence almost orthogonal. This large overlap ensures that the corresponding left edge eigenstates exhibit large amplitudes and contribute most significantly to $P(k)$. Therefore, $P(k)$ is dominated by the occupied edge state contributions:
\bea
P(k)&=&\sum_{m\in occ}\ket{\psi^R_m(k)}\bra{\psi^L_m(k)}\nn\\
&\approx&\ket{\psi^R_{\color{red}{e_1}}(k)}\bra{\psi^L_{\color{red}{e_1}}(k)}=P_\text{edge}(k).
\eea
In the following discussion, for the sake of brevity, we will omit the $k$ in $\ket{\psi^{R(L)}_{e_1}(k)}$.
To proceed, we note that $P_\text{edge}(k)$ is furthermore dominated by just \emph{one} matrix element due to the exponential skin-localization of the edge states. To show this explicitly, we use the similarity transform [Eq~\eqref{H'k}] to write the spatial profiles of the topological SSH edge states in the $y$ direction as
\bea
\ket{\psi'_{e_1}(y)}=\begin{pmatrix}\psi'^A(y)\\ \psi'^B(y) \end{pmatrix}\sim \begin{pmatrix}\epsilon^{y-1}\\\epsilon^{L_y-y}\end{pmatrix}, \nn\\
\text{ where } \epsilon=\left|\frac{t'}{t}\right|=\sqrt{\left|\frac{a_0}{t}(b_0-\cos k)^B\right|}.
\eea
Consequently, the right and left edge states of the original $H_{y-\text{OBC}}$ is given by
\bea
|\psi^R_{\color{red}{e_1}}\rangle=Q|\psi'_{e_1}\rangle\sim\begin{pmatrix}
\begin{pmatrix}1\\ r\epsilon^{L_y-1}\end{pmatrix}\\
r\begin{pmatrix}\epsilon\\ r\epsilon^{L_y-2}\end{pmatrix}\\ \vdots\\
r^{L_y-2}\begin{pmatrix}\epsilon^{L_y-2}\\ r\epsilon\end{pmatrix}\\
r^{L_y-1}\begin{pmatrix}\epsilon^{L_y-1}\\ r\end{pmatrix}
\end{pmatrix}, 
\eea
and
\bea
|\psi^L_{\color{red}{e_1}}\rangle=Q^{-1}\bra{\psi'_{e_1}}\sim
\begin{pmatrix}
\begin{pmatrix}1\\ \epsilon^{L_y-1}/r\end{pmatrix}\\
\frac{1}{r}\begin{pmatrix}\epsilon\\ \epsilon^{L_y-2}/r\end{pmatrix}\\ \vdots\\
\frac{1}{r^{L_y-2}}\begin{pmatrix}\epsilon^{L_y-2}\\ \epsilon/r\end{pmatrix}\\
\frac{1}{r^{L_y-1}}\begin{pmatrix}\epsilon^{L_y-1}\\ 1/r\end{pmatrix}
\end{pmatrix}^T.
\eea
This yields the following form for the projector matrix
\bea
&&P(k)\approx  P_{\text{edge}}(k) = \ket{\psi_{e_1}^R}\bra{\psi_{e_1}^L} \nn \\
&\sim &
\scriptsize
\scalebox{1}{$
\begin{pmatrix}
\begin{pmatrix}1 & \epsilon^{L_y-1}/r\\ r\epsilon^{L_y-1} & \epsilon^{2L_y-2} \end{pmatrix} &
\frac{1}{r}\begin{pmatrix}\epsilon & \epsilon^{L_y-2}/r\\ r\epsilon^{L_y} & \epsilon^{2L_y-3} \end{pmatrix} & \cdots &
\color{blue} \frac{1}{r^{L_y-1}} \begin{pmatrix}\epsilon^{L_y-1} & 1/r\\ r\epsilon^{2L_y-2} & \epsilon^{L_y-1}\end{pmatrix}
\\
r\begin{pmatrix}\epsilon & \epsilon^{L_y}/r\\ r\epsilon^{L_y-2} & \epsilon^{L_y-2}\end{pmatrix} &
\begin{pmatrix}\epsilon^2 & \epsilon^{L_y-1}/r\\ r\epsilon^{L_y-1} & \epsilon^{2L_y-4}\end{pmatrix} & \cdots &
\frac{1}{r^{L_y-2}} \begin{pmatrix}\epsilon^{L_y} & \epsilon/r\\ r\epsilon^{2L_y-3} & \epsilon^{L_y-2}\end{pmatrix}
\\
\vdots & \vdots & \vdots & \vdots \\
r^{L_y-1} \begin{pmatrix}\epsilon^{L_y-1} & \epsilon^{2L_y-2}/r\\ r & \epsilon^{L_y-1}\end{pmatrix} &
r^{L_y-2} \begin{pmatrix}\epsilon^{L_y} & \epsilon^{2L_y-3}/r\\ r\epsilon & \epsilon^{L_y-2}\end{pmatrix} & \cdots &
\begin{pmatrix} \epsilon^{2L_y-2} & \epsilon^{L_y-1}/r\\ r\epsilon^{L_y-1} & 1 \end{pmatrix}
\end{pmatrix}_{2L_y \times 2L_y}
$}
\label{Pedge}
\eea
which is dominated by the upper right matrix element in blue, since $r=\sqrt{\left|\frac{(b_0-\cos k)^B}{a_0}\right|}<1$ for the parameters used.

The truncated projector $\bar{P}$ can be obtained by Fourier transforming each matrix element on the right-hand side (RHS) of Eq~\eqref{Pedge}, i.e.,
\be
\bar{P}=\begin{pmatrix} \bar{P}^{s_1,s_2}_{1,1}& \bar{P}^{s_1,s_2}_{1,2} & \cdots &\bar{P}^{s_1,s_2}_{1,L_y}\\
\bar{P}^{s_1,s_2}_{2,1} & \bar{P}^{s_1,s_2}_{2,2}  & \cdots &\bar{P}^{s_1,s_2}_{2,L_y}\\
\vdots &\vdots &\ddots &\vdots\\
\bar{P}^{s_1,s_2}_{L_y,1} & \bar{P}^{s_1,s_2}_{L_y,2}  & \cdots &\bar{P}^{s_1,s_2}_{L_y,L_y}\\
\end{pmatrix},
\ee
where $\bar{P}^{s_1,s_2}_{y_1,y_2}$ is an $L/2\times L/2$ submatrix with elements
\be
\bra{x_1}{\bar{P}^{s_1,s_2}_{y_1,y_2}}\ket{x_2}=\frac{1}{L}\sum_k e^{ik(x_1-x_2)}[P(k)]^{s_1,s_2}_{y_1,y_2}, 
\ee
with $L/2$ denoting the size of subregion $A$ and $x_1,x_2\in A$. According to Eq~\eqref{Pedge}, the largest element is $[P(k)]^{+,-}_{1,L_y}$, so after Fourier transformation into real space, the dominant submatrix is $\bar P^{+,-}_{1,L_y}$. Substituting $r=\sqrt{\left|\frac{(b_0-\cos k)^B}{a_0}\right|}$, we obtain
\bea
&&\bra{x_1}\bar{P}^{+,-}_{1,L_y}\ket{x_2}=\frac{1}{L}\sum_ke^{ik(x_1-x_2)}[P(k)]^{+,-}_{1,L_y}\nn\\
&\sim & \frac{1}{L}a_0^{L_y/2}\frac{ 2\cos\big(k_1(x_1-x_2)\big)}{(b_0-\cos k_1)^{BL_y/2}}\nn\\
&\sim & a_0^{L_y/2}\Big(\frac{L}{\pi}\Big)^{BL_y-1}\times\Big(1-\frac{1}{2}(\frac{\pi x}{L})^2+\frac{1}{4!}(\frac{\pi x}{L})^4-\frac{1}{6!}(\frac{\pi x}{L})^6+\cdots\Big)\nn\\
&=&c_1L^{BL_y-1}+c_2L^{BL_y-1-2}+\cdots,
\eea
with $b_0=1$ and the leading $k=\pm k_1=\pm\pi/L$ contributions substituted to obtain the 3rd line. The other matrix elements scale more slowly with $L$. At the opposite corner, the submatrix $\bra{x_1}{\bar{P}^{-+}_{L_y,1}}\ket{x_2}\sim O(1)$, so we establish Eq~\eqref{TrP2L}: 
\be
\text{Tr}(\bar{P}^2)\approx 2\text{Tr}(\bar{P}^{+,-}_{1,L_y}\bar{P}^{-,+}_{L_y,1})\sim  c_1L^{BL_y-1}+c_2L^{BL_y-1-2}+\cdots
\ee
where, as a first-order approximation, we have retained only the term containing the largest block $\bar{P}^{+,-}_{1,L_y}$ while disregarded the contributions from other elements. This implies that the eigenvalues $p_i$ generically scale like $p_i\sim cL^{\color{red}\nu}$ with $c$ an unimportant constant. For odd $BL_y$, ${\color{red}\nu}=1,2...,(BL_y-1)/2$ and for even $BL_y$, ${\color{red}\nu}= 1/2, 3/2,...,(BL_y-1)/2$.

For each $p_i=c L^{\color{red}\nu}$ ($c$ is a constant), its contribution to entanglement entropy $S_A$ is
\bea
S_A(p_i)&=&-p_i\log p_i-(1-p_i)\log(1-p_i)\nn\\
&=&- cL^\nu\log(cL^\nu)-(1-cL^\nu)\log(1-cL^\nu)\nn\\
&\approx&-cL^{\nu}\log(cL^\nu)
+cL^\nu\log(-cL^\nu)-\log(1-cL^\nu)\nn\\
&\approx&(cL^\nu+1) \log(-1)-\log(cL^\nu)\nn\\
&\approx& -\nu\log L+cL^\nu\pi i-\log c.
\eea
The other eigenvalue with $p_i'=1-p_i$ (see subsection \hyperref[subsec:C2]{C.2}) contributes the same real part to $S_A$, but opposite imaginary part that cancels off, as numerically observed as twofold degenerate states. Therefore, the total entanglement entropy is dominated by the $\log L$ contribution in blue above which should be multiplied by 4, i.e.,
\bea
S_A&=&\sum_{p_i} -{p_i}\log p_i-(1-p_i)\log(1-p_i)\nn\\
&\sim&
\scriptsize
\scalebox{1.2}{$
-4\times\left[\frac{BL_y-1}{2}+\left(\frac{BL_y-1}{2}-1\right)+\left(\frac{BL_y-1}{2}-2\right)+\cdots\right]\log L.
$}
\label{eq28}
\eea
Calling $J=\frac{BL_y-1}{2}$ ($B,L_y\in \mathbb{Z}$) and $J'=\lceil J\rceil=\begin{cases}J,& BL_y\text{ is odd}\\ J+1/2, & BL_y\text{ is even}\end{cases}$, we finally obtain the scaling relation of $S_A$ with $L,L_y$ as
\bea
S_A&\sim& -4\log L\Big(J+(J-1)+(J-2)+\cdots+(J-J'+1)\Big)\nn\\
&=&-2J'(2J-J'+1)\log L\nn\\
&=&-\frac{1}{2}\Big((BL_y)^2-(BL_y\text{mod} 2)\Big)\log L \, \nn\\
&=& -\left\lceil\frac{B^2L_y^2-1}{2}\right\rceil\log L\quad\sim-\frac{1}{2}(BL_y)^2\log L, 
\label{eq29}
\eea
which is our key result: an unconventional quadratic negative entanglement scaling.

\begin{figure*}
\centering
\includegraphics[width=15cm, height=10cm]{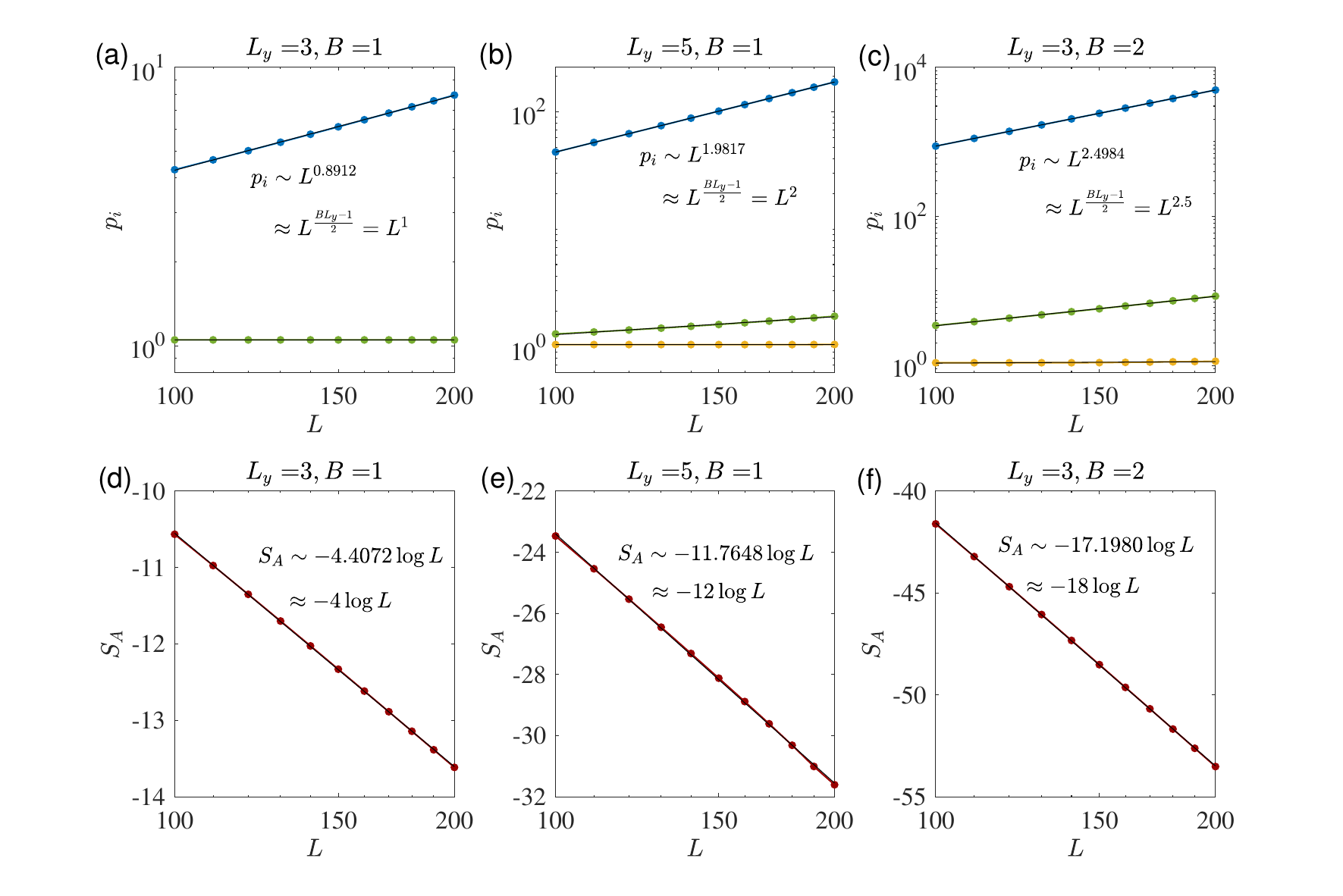}
\caption{Excellent agreement of our quadratic negative entanglement scaling result with numerics for various $B$ and $L_y$, both for the $\bar P$ eigenvalues $p_i$ [Eq~\eqref{TrP2L}] (a-c) and the corresponding entanglement scaling $S_A\sim -\left\lceil(B^2L_y^2-1)/2\right\rceil\log L$ [Eq~\eqref{eq29}]. Parameters: $t=0.5, a_0=2, b_0=1$.}
\label{FigS4}
\end{figure*}

Below, we further present the numerical verification of Eq~\eqref{eq28}. In Fig.~\ref{FigS4}(a), we take $L_y=3, B=1$, since $\frac{BL_y-1}{2}=1$, only the first order $p_i\sim L^1$ exists. The numerical fitting shown in Fig.~\ref{FigS4}(d) yields:
\be S_A\sim-4.4072\log L\quad\approx-4\times\frac{BL_y-1}{2}\log L=-4\log L,\ee
which demonstrates good agreement with Eq~\eqref{eq28}. In Fig.~\ref{FigS4}(b), where $L_y=5, B=1$, $\frac{BL_y-1}{2}=2$, both the first and second orders of $L$ in $p_i\sim L^1$, $p_i\sim L^2$ are present. We numerically obtain the scaling of entanglement entropy as depicted in Fig.~\ref{FigS4}(e):
\bea
&S_A& \sim -11.7648\log L \nn\\
&\approx& -4\times\left(\frac{BL_y-1}{2}+\frac{BL_y-1}{2}-1 \right)\log L=-12\log L.
\eea
For the case of $L_y=3, B=2$ shown in Fig.~\ref{FigS4}(c), the first, second and third orders of $L$ all contribute to the EE:
\bea
S_A &\sim& -17.1980\log L\nn\\
&\approx & -4\times\left(\frac{BL_y-1}{2}+\frac{BL_y-1}{2}-1+\frac{BL_y-1}{2}-2\right)\log L\nn\\
&=&-18\log L.
\eea
While plots with even larger $L_y$ and $B$ would be more prone to numerical errors due to the larger $L$ needed, excellent agreement is already observed for this cases which corresponds to the most realistic models.

\begin{figure*}
\centering
\includegraphics[width=18cm, height=5cm]{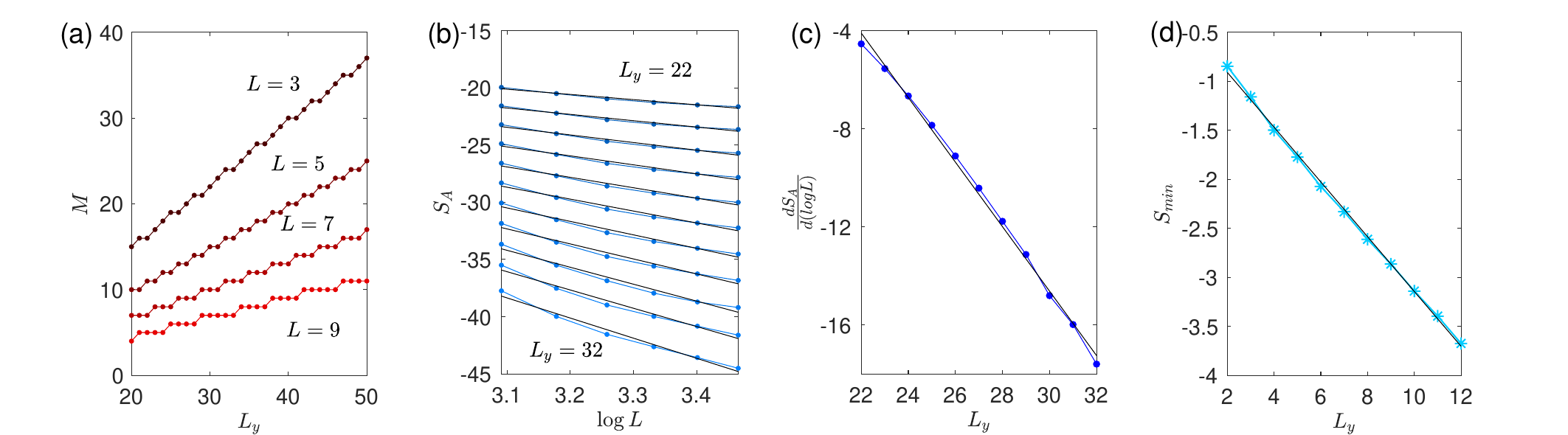}
\caption{Negative entanglement scaling for the gapped EP case ($b_0>1$). (a) As numerically plotted for $L=9, b_0=1.2$, the approximate linear scaling of the scaling exponent $M$ of the dominant $\bar P$ eigenvalue with $L_y$, as described by Eq~\eqref{S48}.  (b) The negatively linear relationship of the entanglement entropy $S_A$ with $\log L$ for different $L_y $ ranging from 22 to 32, with $L\in[22,32]$ and parameters $t=0.8,a_0=1, b_0=1.2, B=1$. As predicted in Eq~\eqref{S51}, the gradient indeed increases with $L_y$.  
 (c) Using the same parameters as in (b), approximate linear dependence of the gradient of $S_A$ on $L_y$. (d) For large $L=60$ and parameters $t=0.8, B=1, b_0=1.2, a_0=2$, $S_{min}$ indeed scales linearly with $L_y$, as predicted in Eq~\eqref{Smin}.}
\label{FigS5}
\end{figure*}

\subsubsection{ Scaling of entanglement entropy \texorpdfstring{$S_A$}{SA} for gapped flat-band edge states}
\label{subsec:B3}
In the case of $b_0>1$ (e.g., $b_0=1.2$), there exists a small gap $\Delta$ in the edge bands. Due to this gap, the entanglement entropy $S_A$  decreases with increasing system size $L$ and eventually stabilizes at a minimum constant value, $S_{min}$. In this section, we will demonstrate that: 
\\
1) For $L\ll L_y$, $S_A$ exhibits an approximate negative $L_y$-linear scaling with $\log L$, expressed as $S_A\sim -({\color{red}\kappa} L_y+{\color{red}\xi})\log L$, where $\kappa$ and $\xi$ are fitting coefficients with $\alpha>0$; \\
2) While for $L\gg L_y$, $S_A$ saturates to the lower bound, $S_{min}$, which scales linearly with $L_y$ as: $S_{min}\sim -L_y\log\left(\frac{a_0}{(b_0-1)^B}\right)$.

As discussed above, we demonstrated that near $k=0$, the dominant element of $P(k)$ is located in the upper right corner, where $[P(k)]^{+,-}_{1,L_y}\sim 1/r^{L_y}$ with $r=\sqrt{\big|\frac{(b_0-\cos k)^B}{a_0}\big|}<1$. For $b_0>1$, this can still be the case with appropriate choice of $a_0$. We have
\bea
\bra{x_1}\bar{P}^{+,-}_{1,L_y}\ket{x_2}&=&\frac{1}{L}\sum_ke^{ik(x_1-x_2)}[P(k)]^{+,-}_{1,L_y}\nn\\
&\sim& \frac{1}{L}\sum_ke^{ik(x_1-x_2)}
a_0^{L_y/2}\frac{1}{(b_0-\cos k)^{BL_y/2}},
\label{S46}
\eea
where $k$ takes the values $\pi/L, 3\pi/L, 5\pi/L,...$. 
For $k_1=\pi/L$, which contributes most significantly to $\bra{x_1}\bar{P}^{+-}_{1,L_y}\ket{x_2}$, we have
\bea
(b_0-\cos k_1)^{BL_y/2}&=&(b_0-1+1-\cos k_1)^{BL_y/2}\nn\\
&=&(b_0-1)^{BL_y/2}(1+x)^{BL_y/2},
\label{S47}
\eea
where $x=\frac{1-\cos k_1}{b_0-1}\approx \frac{\pi^2}{2(b_0-1)L^2}$.

When $L\ll L_y$, (for simplicity, we select $L_y$ to be an even number to ensure $BL_y/2$ is an integer), the following approximation holds 
\be
(1+x)^{BL_y}=\sum_m C_{BL_y/2}^m x^m \approx C_{BL_y/2}^M x^M
\label{S48}
\ee
where $m=M$ represents the maximum term in the summation. As verified in Fig.~\ref{FigS5}(a), $M$ exhibits a linear relation with $L_y$, expressed as $M\approx\frac{1}{2}(\gamma L_y+\zeta)$ (where $\gamma, \zeta$ are coefficients independent of $L_y$). Consequently, we find 
\bea
(1+x)^{BL_y}&\approx & C_{BL_y/2}^M x^M\sim C_{BL_y/2}^{(\gamma L_y+\zeta)/2}\left(\frac{\pi^2}{2(b_0-1)L^2}\right)^{(\gamma L_y+\zeta)/2}\nn\\
&\sim & L^{-(\gamma L_y+\zeta)}.
\eea
By substituting into Eqs~\eqref{S47} and \eqref{S46} and considering only the contribution from $k_1=\pi/L$, we have
\bea
\bra{x_1}\bar{P}^{+,-}_{1,L_y}\ket{x_2}&\sim &\frac{1}{L}e^{ik_0(x_1-x_2)}\frac{a_0^{L_y/2}}{(b_0-1)^{BL_y/2}(1+x)^{BL_y/2}}\nn\\
&\sim& L^{\gamma L_y+\zeta-1}.
\eea
As we have demonstrated in the previous subsection \hyperref[subsec:B2]{B.2}, since the elements of $\bar{P}^{+,-}_{1,L_y}$ are proportional to $L^{\gamma L_y+\zeta-1}$, we can deduce:
\bea
\text{Tr}(\bar{P}^2)&\approx &2\text{Tr}(\bar{P}^{+,-}_{1,L_y}\bar{P}^{-,+}_{L_y,1})\sim L^{\gamma L_y+\zeta-1}\nn\\
\Longrightarrow p_i^{1st}&\sim &L^{\frac{\gamma L_y+\zeta-1}{2}},\nn\\
S_A&\sim& -4\times\left(\frac{\gamma L_y+\zeta-1}{2}\right)\log L\nn\\
&\sim& -({\kappa} L_y+{\xi})\log L,
\label{S51}
\eea
where we have redefined $2\gamma$ as $\kappa$ and $2(\zeta-1)$ as $\xi$. In the $L\ll L_y$ regime, higher orders of $p_i$ are negligible, and we only need to consider the first order.  Therefore, we conclude: for $L\ll L_y$, $S_A\sim -\log L$ (Fig.~\ref{FigS5}(b)), with the gradient depending linearly on $L_y$ (see Fig.~\ref{FigS5}(c)). Even though our derivation had assumed $L\ll L_y$, from the numerics in Fig.~\ref{b0ne1Ek}(e) of the main text, we see that this trend still mostly holds as long as $L<L_y$.

When $L\gg L_y$, Eq~\eqref{S47} can be approximated as:
\bea
(b_0-\cos k_0)^{BL_y/2}&=&\left(b_0-1+\frac{1}{2!}\left(\frac{\pi}{L}\right)^2-\frac{1}{4!}\left(\frac{\pi}{L}\right)^4+\cdots\right)^{BL_y/2}\nn\\
&\approx&(b_0-1)^{BL_y/2}.
\eea
Thus, Eq~\eqref{S46}, which represents the elements of the largest matrix block $\bar{P}^{+,-}_{1,L_y}$, can be expressed as
\be
\bra{x_1}\bar{P}^{+,-}_{1,L_y}\ket{x_2}\sim \left(\frac{a_0}{(b_0-1)^B}\right)^{L_y/2}.
\ee
Following a similar derivation as Eq~\eqref{S51}, we now have 
\bea
\text{Tr}(\bar{P}^2)&\sim &\left(\frac{a_0}{(b_0-1)^B}\right)^{L_y/2}, \quad p_i^{1st}\sim\left(\frac{a_0}{(b_0-1)^B}\right)^{L_y/4}\nn\\
S_{min}&\sim& -4\times \frac{L_y}{4}\log\left(\frac{a_0}{(b_0-1)^B}\right)\nn\\
&=&-L_y\log\left(\frac{a_0}{(b_0-1)^B}\right).
\label{Smin}
\eea
This linear dependence on $L_y$ is verified in Fig.~\ref{FigS5}(d). Empirically, we also see from Fig.~\ref{b0ne1Ek}(e) of the main text that this saturation generally sets in as long as $L$ exceeds $L_y$. 

\begin{figure*}
\centering
\includegraphics[width=17cm, height=5cm]{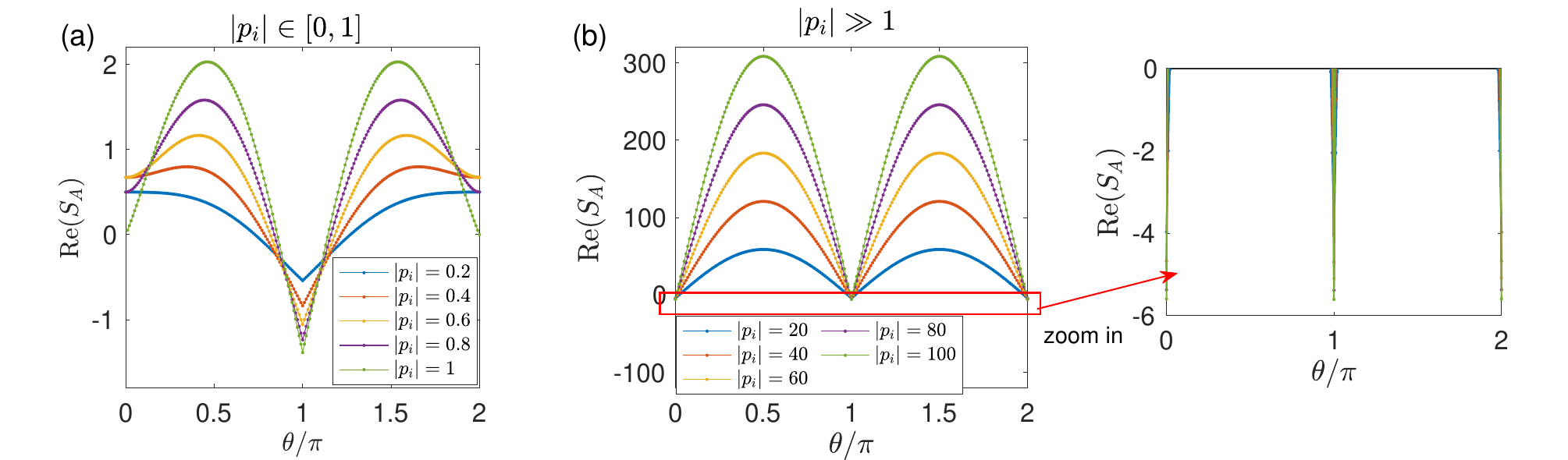}
\caption{ Contribution of a complex $p_i=|p_i|e^{i\theta}$ to the real part of $S_A$ for two cases: (a) $|p_i|\in[0,1]$ and (b) $|p_i|\gg 1$.}
\label{FigS6}
\end{figure*}

\begin{figure*}
\centering
\includegraphics[width=15cm, height=9cm]{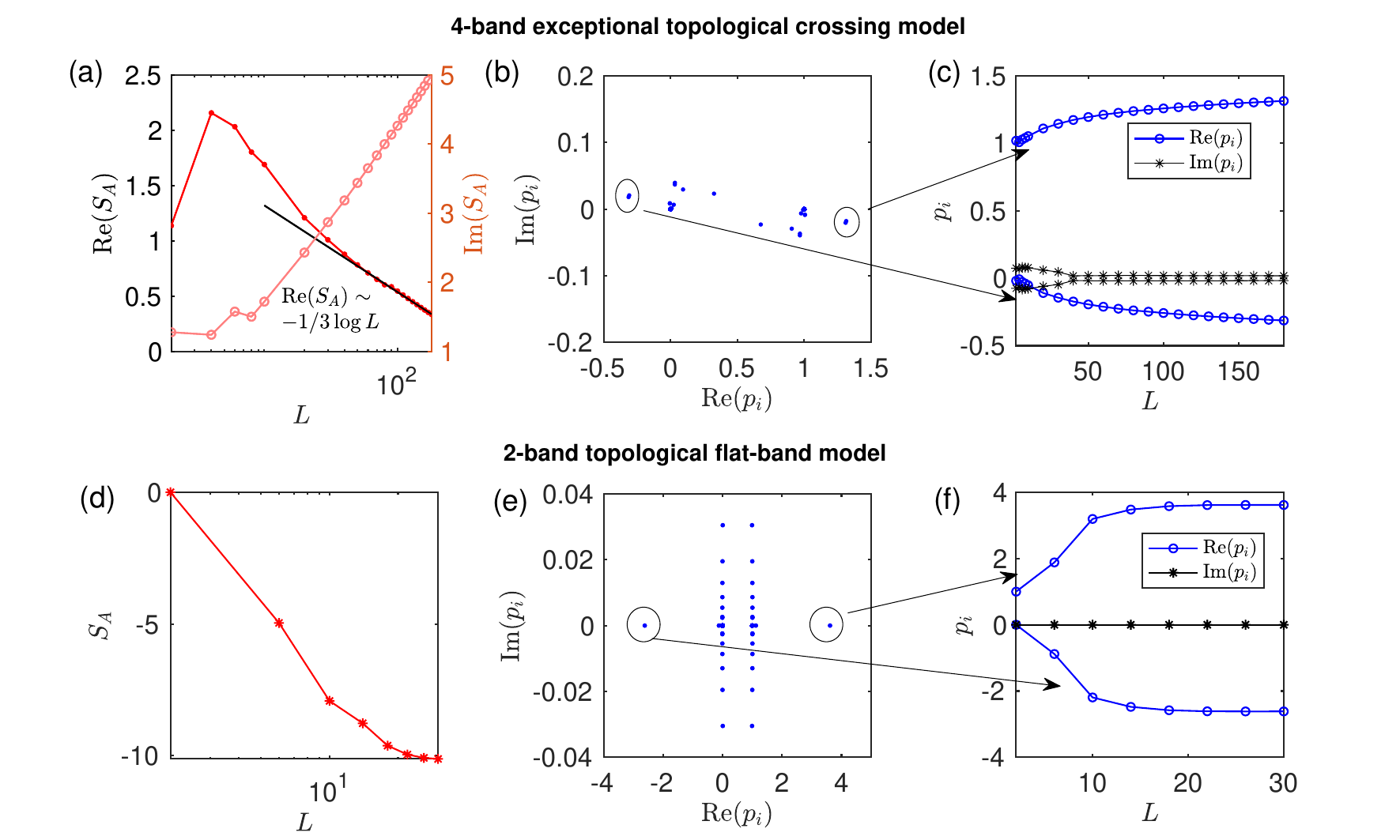}
\caption{Negligible imaginary part for complex $p_i$ with Re$(p_i)>1$ or Re$(p_i)<0$ in both the 4-band exceptional crossing model (a-c) and the 2-band flat-edge-band model (d-f). (a) Replot of the topo, non-Hermitian case from Fig. 1(d). (c) Corresponding scaling of Re($p_i$) and Im($p_i$) for the complex $p_i$ circled in (b).
(d) Replot of the $L_y=15$ case from Fig. 2(e). (f) Corresponding scaling of Re($p_i$) and Im($p_i$) for the complex $p_i$ circled in (e).
}
\label{FigS7}
\end{figure*}

\subsection{Further properties of the negative entanglement}

\subsubsection{Complex $p_i$ and the negativity of von Neumann and R\'enyi entropy}
\label{subsec:C1}
In general non-Hermitian systems, the eigenvalues $p_i$ of $\bar{P}$ are complex, making it unintuitive in determining how $p_i$ affects the real part of $S_A$. In this section, we discuss this issue in detail. 
Most importantly, we will demonstrate that the negativity in entanglement is not only limited to the von Neumann entropy, but also can be extended to nth-order ($n\ge2 $) R\'enyi entropies.

First, for the von Neumann entropy, we consider a general complex eigenvalue $p_i=|p_i|e^{i\theta}$, analyzing two cases: $|p_i|\in[0,1]$ and $|p_i|\gg 1$. As shown in Fig. \ref{FigS6}(a), for $|p_i|\in[0,1]$, Re($S_A$) becomes negative near $\theta=\pi$, where Re$(p_i)<0$ and Im($p_i)$ is approximately zero. For the case of $|p_i|\gg 1$, the negative contribution occurs more sharply around $\theta=0,\pi$ (see the inset of Fig. \ref{FigS6}(b)), where Re$(p_i)$ is either much less than 0 or greater than 1, while Im($p_i)$ remains negligible.

Thus, for the von Neumann entropy, we conclude that for nearly real $p_i$, when Re$(p_i)>1$ or Re$(p_i)<0$, it contributes negatively to Re$(S_A)$. This conclusion can also be derived as follows:
\bea
&&\text{Re}[S_A(p_i)]=\text{Re}[-p_i\log(p_i)-(1-p_i)\log(1-p_i])\nn\\
&=&\begin{cases}
-p_i\log(p_i)+(p_i-1)\log(p_i-1)<0,\text{ real } p_i >1\\
|p_i|\log(|p_i|)-(|p_i|+1)\log(|p_i|+1)<0,\text{ real }p_i<0
\end{cases}
\eea
Since the above discussion is based on the condition that $p_i$ is nearly real, we numerically calculated the values of $p_i$ for the two models introduced in our paper. As shown in Figs.~\ref{FigS7}(c) and (f), we focus primarily on the $p_i$ values whose real parts extend beyond the range $[0,1]$ (as highlighted by the circles in Figs.~\ref{FigS7} (b) and (e)). We observe that Im$(p_i)$ remains close to zero, indicating that it is indeed negligible compared to Re$(p_i)$. 
As expected, these $p_i$ contribute to a negative scaling of Re$(S_A)$ for both models, as shown in Figs.~\ref{FigS7} (a) and (d).

Secondly, we derive how the complex $p_i$ contributes to the real part the $n$-th order R\'{e}nyi entropy. For the case $|p_i|\gg 1$, which, as demonstrated in the main article, is induced by the large overlap of different eigenstates, we have:
\bea
p_i^n+(1-p_i)^n\sim \begin{cases}
p_i^n + O(p_i^{n-1}),& \text{for even $n$}\\
p_i^{n-1} +O(p_i^{n-2}),& \text{for odd $n$}
\end{cases}
\eea
Thus, considering only the dominant term, we approximately have:
\bea
S_A^{(n)}(p_i)&=&\frac{1}{1-n}\log(p_i^n+(1-p_i)^n)\nn\\
&\sim&\begin{cases}
\frac{n}{1-n}(\log|p_i|+i\arg (p_i)), & \text{for even $n$}\\
-(\log|p_i|+i\arg (p_i)),& \text{for odd $n$}
\end{cases},\nn\\
\text{Re}[S_A^{(n)}(p_i)]&\sim &\begin{cases}
\frac{n}{1-n}\log|p_i|, & \text{for even $n$}\\
-\log|p_i|,& \text{for odd $n$}
\end{cases} <0.
\eea
Hence, unlike the von Neumann entropy, for general complex $p_i$ with $|p_i|\gg 1$, even if the imaginary part is not negligible, its contribution to the real part of R\'{e}nyi entropy is always negative.

\subsubsection{PT-symmetry-protected real entanglement entropy \texorpdfstring{$S_A$}{SA} in the flat band model.} 
\label{subsec:C2}
In this section, we demonstrate that both the von Neumann entropy and R\'{e}nyi entropies are rigorously real for PT-symmetry-protected models, as exemplified by our flat-edge-band model. This result arises from the fact that the eigenvalues $p_i$ of $\bar P$ consistently occur in complex conjugate pairs.

First, in the $H_\text{y-OBC}(k)$ model of Eq~\eqref{HsshzOBC}, the parameters $a_0,b_0,t$ are always real numbers, with
$b_0\ge 1$. As previously demonstrated, this model can be viewed as a non-Hermitian SSH model that is protected by PT-symmetry, and it features purely real eigenvalues and eigenvectors, $|\psi_m^R(k)\rangle$ and $\langle\psi_m^L(k)|$.
Therefore, the elements of the projector matrix ${\left[P(k)\right]_{y_1,y_2}^{s_1,s_2}=\sum_{m\in\text{occ}}\langle y_1,s_1|\psi_m^R(k)\rangle\langle\psi_m^L(k)|y_2,s_2\rangle}$ are all real, with $s_1,s_2$ labeling the sublattice indices. Upon expanding these elements into real space, and given that $\left[P(k)\right]_{y_1,y_2}^{s_1,s_2}$ are real, we will obtain
\bea
\bra{x_1,y_1,s_1}P\ket{x_2,y_2,s_2}&=&\sum_k \left[P(k)\right]_{y_1,y_2}^{s_1,s_2}e^{ik(x_1-x_2)}\nn\\
&=&\Big(\langle x_2,y_2,s_2|P|x_1,y_,s_1\rangle\Big)^*.
\eea
Hence, the real-space truncated projector $\bar P$ satisfy 
\be
V\bar PV^{-1}=\text{conj}(\bar P), \text{ with } V=\begin{pmatrix}0&0&0&\cdots &\mathbb{I}\\0&0&\cdots&\mathbb{I}&0\\0&\cdots&\mathbb{I}&0&0\\ \vdots&\vdots&\vdots&\vdots&\vdots \\ \mathbb{I}&0&0&\cdots&0
\end{pmatrix},\ee
indicating that $\bar P$ possesses PT-symmetry and its spectrum must consists of complex conjugate pairs $(p,p^*)$.

For the conjugate pairs $(p,p^*)=\left(|p|e^{i\theta},|p|e^{-i\theta}\right)$, their contribution to the von Neumann entropy is:
\bea
S_A^{\text{pair}}&=&-|p|e^{i\theta}\log(|p|e^{i\theta})-(1-|p|e^{i\theta})\log (1-|p|e^{i\theta})\nn\\
&&-|p|e^{-i\theta}\log(|p|e^{-i\theta})-(1-|p|e^{-i\theta})\log (1-|p|e^{-i\theta})\nn\\
&=&
\scriptsize
\scalebox{1.3}{$
2|p|\left(\theta\sin\theta-\log |p|\cos\theta\right)+2|p'|(\theta'\sin\theta'-\log |p'|\cos\theta'),
$}\nn\\
\eea
where $|p'|,\theta'$ are defined such that $1-|p|e^{i\theta}=|p'|e^{i\theta'}$. As shown above, $\Im(S_\text{pair})=0$, indicating that the von Neumann entropy should be a purely real. And their contribution to the $n-$th order R\'enyi entropy is:
\bea
\left(S_A^{(n)}\right)^{\text{pair}}&=&\frac{\log\left[p^n+(1-p)^n\right]}{1-n}+\frac{\log\left[(p^*)^n+(1-p^*)^n\right]}{1-n}\nn\\
&=&\frac{1}{1-n}\log(|p''|e^{i\theta''})+\frac{1}{1-n}\log(|p''|e^{-i\theta''})\nn\\
&=&\frac{2}{1-n}\log|p''|,
\eea
where we have defined 
\be
p^n+(1-p)^n=|p''|e^{i\theta''}.
\ee
This result shows that  $\Im\left[\left(S_A^{(n)}\right)^{\text{pair}}\right]=0$, i.e., the R\'enyi entropies are purely real. Therefore, any observed imaginary part of the entanglement entropy in numerical computations must be attributed to numerical errors.

\subsubsection{Further discussion on the role of  $\bar{P}$ eigenvalues. }
In many-body free fermion systems, as discussed in this paper, the von Neumann entropy (or R\'enyi entropy) is a mathematical function of the eigenvalues $p_i$ of the truncated projector operator $\bar{P}$. This highlights that the truncated projector operator itself can be regarded as a more fundamental quantity, encapsulating the essential features of the system's entanglement structure.

As highlighted in previous studies \cite{EBstate}, the presence of significant coalescence between the right eigenstates causes the momentum-space projector $P(k)$ to approach singularity. When transformed into real space, this results in the projector operators $P$ and $\bar{P}$ exhibiting slowly decaying, long-range matrix hopping elements. 
In contrast, in trivial cases with no substantial overlap, $\bar{P}$ is characterized by short-range matrix hopping elements. A striking consequence of these long-range matrix hopping elements is the emergence of unique eigenmodes, referred to as Exceptional Bound States. These modes are distinguished by the real parts of their eigenvalues extending well beyond the typical range of $[0,1]$, ultimately contributing to a negative real component of the entanglement entropy (EE). 
Thus, the emergence of these modes represents a fundamental and profound consequence of the large overlap.

Hence, even without focusing on the entanglement entropy, the presence of a large eigenstate overlap can be regarded as a physical phenomenon in itself if the $\bar P$ matrix elements are taken to be physical hoppings. For instance, by constructing a long-range hopping system i.e. a circuit metamaterial array and designing the Hamiltonian to mimic the topologically protected, non-trivial $\bar{P}$ introduced in this paper, one would observe a novel energy spectrum. This spectrum features special isolated eigenenergies that lie well outside the conventional range of $[0,1]$.

\bibliography{topoEB}

\noindent{\bf Acknowledgments} 

\noindent{We thank Ruizhe Shen and Linhu Li for helpful discussions. This work is supported by the National Research Foundation, Singapore under its QEP2.0 programme (NRF2021-QEP2-02-P09) as well as the Ministry of Education, Singapore (MOE award number: MOE-T2EP50222-0003).

\end{document}